\documentclass[journal]{IEEEtran}
\ifCLASSINFOpdf
\else
  \usepackage[dvipdf]{graphicx}
\fi
 \usepackage{amssymb}
\usepackage{amsmath}
\usepackage{diagbox}
\usepackage{graphicx}
\usepackage{epstopdf}
\usepackage{booktabs}
\usepackage[justification=centering]{caption}

\usepackage{multirow}
\usepackage{slashbox}
\usepackage{color}
\usepackage{colortbl}
\usepackage{makecell}
\usepackage{algorithm}
\usepackage{algorithmic}

\usepackage{subfigure}
\usepackage{multicol}

 \renewcommand{\algorithmicrequire}{\textbf{Input:}} 
\renewcommand{\algorithmicensure}{\textbf{Output:}} 
\begin{document}
%
\title{ Movable Antenna Empowered Physical Layer Security Without Eve's CSI: Joint Optimization of Beamforming and Antenna Positions }

%
%
%
\author{Zhiyong Feng,~\IEEEmembership{Senior Member,~IEEE}, Yujia Zhao, Kan Yu,~\IEEEmembership{Member,~IEEE}, and Dong Li,~\IEEEmembership{Senior Member,~IEEE}
\thanks{This work is supported by the National Natural Science Foundation of China with Grants 62301076, the Macao Young Scholars Program with Grants AM2023015, the Natural Science Foundation of Shandong Province with Grants ZR2021QF050 and ZR2021MF075, and the Science and Technology Development Fund, Macau SAR, under Grant 0029/2021/AGJ.
}
\thanks{Z. Feng is with the Key Laboratory of Universal Wireless Communications, Ministry of Education, Beijing University of Posts and Telecommunications, Beijing, 100876, P.R. China. E-mail: fengzy@bupt.edu.cn;}
\thanks{Y. Zhao is with the Key Laboratory of Universal Wireless Communications, Ministry of Education, Beijing University of Posts and Telecommunications, Beijing, 100876, P.R. China. E-mail: zhaoyj@bupt.edu.cn;}
\thanks{K. Yu (corresponding author) is with 
the Key Laboratory of Universal Wireless Communications, Ministry of Education, Beijing University of Posts and Telecommunications, Beijing, 100876, P.R. China; the School of Computer Science and Engineering, Macau University of Science and Technology, Taipa, Macau, 999078, P. R. China. E-mail: kanyu1108@126.com.}
\thanks{D. Li (corresponding author) is with School of Computer Science and Engineering, Macau University of Science and Technology, Taipa, Macau, China. E-mail: dli@must.edu.mo.}


}

%
%

\markboth{IEEE Transactions on Wireless Communications,~Vol.~, No.~, 2024}%
{Shell \Baogui Huang{\textit{et al.}}: Shortest Link Scheduling Under SINR}
%



\maketitle

\begin{abstract}
Physical layer security (PLS) technology based on the fixed-position antenna (FPA) has {attracted widespread attention}. Due to the fixed feature of the antennas, current FPA-based PLS schemes cannot fully utilize the spatial degree of freedom, and thus a weaken secure gain in the desired/undesired direction may exist. Different from the concept of FPA, mobile antenna (MA) is a novel technology that {reconfigures} the wireless channels and enhances the corresponding capacity through the flexible movement of antennas on a minor scale. MA-empowered PLS enjoys huge potential and deserves further investigation. In this paper, we, for the first time, investigate the secrecy performance of MA-enabled PLS system where a MA-based Alice transmits the confidential information to multiple single-antenna Bobs, in the presence of the single-antenna eavesdropper (Eve) {in the absence} of perfect channel state information (CSI). For the purpose of the secrecy rate maximization of the worst Bob, we jointly design the transmit beamforming and antenna positions at the Alice, subject to the minimum moving distance of the antenna, uncertainty CSI of Eve, and maximum transmit power. Furthermore, the projected gradient ascent (PGA), alternating optimization (AO), and simulated annealing (SA) {are} adopted to solve the non-convex characteristics of the problem of the secrecy rate maximization. Simulation results demonstrate the effectiveness and correctness of the proposed method. In particular, MA-enabled PLS scheme can significantly enhance the secrecy rate compared to the conventional FPA-based ones for different settings of key system parameters.

\end{abstract}


%
\IEEEpeerreviewmaketitle

\section{Introduction}\label{intro}
The Internet of Vehicles (IoV) is an emerging paradigm that leverages wireless communication technologies to enable seamless connectivity and communication {among} vehicles, infrastructure, and other smart devices \cite{ref0.001}. Due to the {broadcasting} nature of wireless channels and characteristics of high-speed vehicle movement, there is a pressing need for innovative solutions to address security challenges.
Physical layer security (PLS), a critical aspect of IoV security, focuses on protecting wireless communication channels from eavesdropping, interception, and other malicious activities. 
Unlike traditional cryptographic methods that rely on {the computationally complex} data encryption and decryption, PLS leverages the inherent randomness and unpredictability of wireless channels to ensure secure communication between devices, without the need for complex cryptographic algorithms. {In addition}, there {exist} many popular PLS schemes, such as the reconfigurable intelligent surface (RIS)  \cite{ref0.002,ref0.003}, artificial noise (AN) \cite{ref0.004,ref0.005}, and beamforming  \cite{ref0.006,ref0.007}.

In essence, the aforementioned methods are based on the fixed-position antenna (FPA), where all antennas are deployed at fixed positions and corresponding steering vector is static for a given fixed steering angle. As a result, the schemes of AN, RIS and beamforming cannot fully utilize the spatial degree of freedom (DoF), leading to a weaken secure gain in the desired/undesired direction. To overcome this limitation, movable antenna (MA) technology has been proposed and demonstrated the great potential to secure wireless communications \cite{ref0.008}. Unlike FPA, MAs offer the flexibility to dynamically adjust their orientation and configuration, allowing for optimal signal transmission and reception in dynamic environments by varying the steering
vectors corresponding to different angles, thereby enhancing the secrecy performance of IoV systems.

{Currently, based on the fact that the movement of antennas enables more flexible beamforming, there are two most related studies on the MA-enabled PLS \cite{ref0.008,ref0.009}. In \cite{ref0.008}, Cheng \emph{et al.} considered the scenario where a multi-antenna transmitter communicates with a single-antenna receiver in the presence of an eavesdropper (Eve), and enhanced the security of MA-enabled wireless systems by jointly optimizing the beamformer and antenna positions of the transmitter, under the constraints of power consumption and secrecy rate. 
While in \cite{ref0.009}, Hu \emph{et al.} further investigated the MA-enabled PLS systems for the scenario in the presence of multiple single-antenna and colluding Eves. The secrecy rate can be maximized by jointly designing the transmit beamforming and positions of all antennas at the transmitter, under the constraints of tranmit power budget and specified regions for positions of all MAs.
In the previous works, to reveal  the fundamental secrecy rate limit of the MA-enabled or FPA-enabled secure communication system, both the perfect channel state information (CSI) of legitimate users and Eves usually was assumed to be available at the transmitter \cite{ref0.008,ref0.009}.
However, the perfect assumption of Bobs and Eves' CSI remains a significant challenge in a IoV system \cite{ref0.0010}. On the one hand, due to the vehicles movement, the obtained CSI may be outdated. On the other hand, Eves can adopt passive eavesdropping means to disguise the positions of themselves.}

{Different from all previous works on the study of MA-enabled PLS systems, considering the scenario where a MA-enabled Alice transmits information to several single-antenna Bobs, in the presence of a single-antenna Eve without position information, in this paper, we maximize the secrecy performance of the MA-assisted PLS system by joint optimization of the beamforming and antenna positions of MAs at the transmitter with no CSI of the Eve. The main contributions of this paper are summarized as follows.
\begin{itemize}
    \item First, we establish a framework of modeling Eve's imperfect CSI based on the concept of a virtual MA. To the best of the author's knowledge, this is the first work that takes into account CSI uncertainty for the physical layer security enabled by the MA; 
    \item Furthermore, considering the worst case where the Eve can move to the optimal eavesdropping position, we aim to maximize the secrecy rate of the Bob with the worst channel quality by jointly optimize the positions and beamforming matrix of MAs at Alice. The projected gradient ascent (PGA) method and alternating optimization (AO) method are employed to effectively solve {the} non-convex problem. In addition, solely using methods of PGA and AO may lead to the optimization process to get stuck in a local optimal solution or even does not reach the convergence, the simulated annealing (SA) is introduced to find a global optimum and ensure the feasibility and efficiency of the solution;
    \item Finally, numerical results demonstrate the effectiveness, feasibility, and convergence of the proposed method by jointly designing beamforming matrix and antenna positions, and shows that MA-enabled PLS can significantly enhances the secrecy rate compared to the FPA-based ones, {where} the spatial DoF of MA can be fully exploited.
\end{itemize}}

{The rest of the paper is organized as follows. In Section \ref{sec:network model}, we introduce the network model and conduct a comprehensive problem analysis, along with presenting the necessary preliminaries. Section \ref{sec:strategies} is dedicated to presenting our proposed methods for solving the optimization problem via the SA. we describe the experiments in Section \ref{sec:evaluations}. Finally, Section \ref{sec:conclusion} concludes the paper and summarize our findings.}

{\emph{Notations:} In this paper, $\textbf{X}^T$ and $\textbf{X}^H$ stand for the transpose and conjugate transpose of matrix $\textbf{X}$, respectively. \(x^*\) indicates the conjugate of the complex number \(x\). $ \mathbb{C}^{a \times b}$ and $ \mathbb{R}^{a \times b}$ respectively denote \(a \times b\) dimensional complex matrices and \(a \times b\) dimensional real matrices. \(\text{tr}(\mathbf{X})\) and \(\rm{diag}(x)\) respectively represent the trace and the diagonal matrix with diagonal elements  \(\mathbf{X}\). \(\mathcal{CN}(a, b)\) represents a complex Gaussian distribution with mean  \(a\) and variance  \(b\). \(\left| \mathbf{x} \right|\) represents taking the modulus of vector \(\mathbf{x}\). \(\nabla_{x}\) and \(\frac\partial{\partial x}\) respectively denote the gradient operator and the partial derivative operator. \([x]^{+} \text{represents} \max\{x, 0\}\). }{Multivariable vector-valued function $\mathbf{f}(\mathbf{X})$ {with} $\mathbf{X} = [\mathbf{x}_1, \dots ,\mathbf{x}_N]$ and  $\mathbf{x}_{n'}$ {denotes} the collection of vectors in $\mathbf{X}$ that does not include $\mathbf{x}_n$. Then $\mathbf{f} \left(\mathbf{x}_n;\mathbf{x}_{n'}\right)$, {that} $\mathbf{f}$ is a function of {the} only variable $\mathbf{x}_n$, while treating other variables $\mathbf{x}_{n'}$ as constants.}

\section{System Model And Problem Formulation}\label{sec:network model}
{In this section, we first present the system model for secure MA-assisted multiuser downlink wireless communication. Due to the covertness of Eve, the corresponding perfect CSI can hardly be obtained. Accordingly, we then model the imperfect CSI and randomness of Eve with the concept of MA, i.e., utilizing a MA, denoted as the virtual MA, to model the set of potential locations of the Eve. To the best of the {authors'} knowledge, it is the first {attempt} to {investigate} the imperfect CSI and randomness of Eve, which is more practical than {the previous works}, since {the} imperfect CSI model of Eve based on virtual MA can construct a more accurate channel matrix and provide more detailed Eve's information. Finally, we construct an optimization problem that ensures the perfect secrecy of the system by jointly optimizing the beamforming of {the} legitimate user and the positioning of MA.}

\subsection{MA-Assisted System Model}
\begin{figure}[htbp]
  \centering 
  \includegraphics[width=0.32\textwidth]{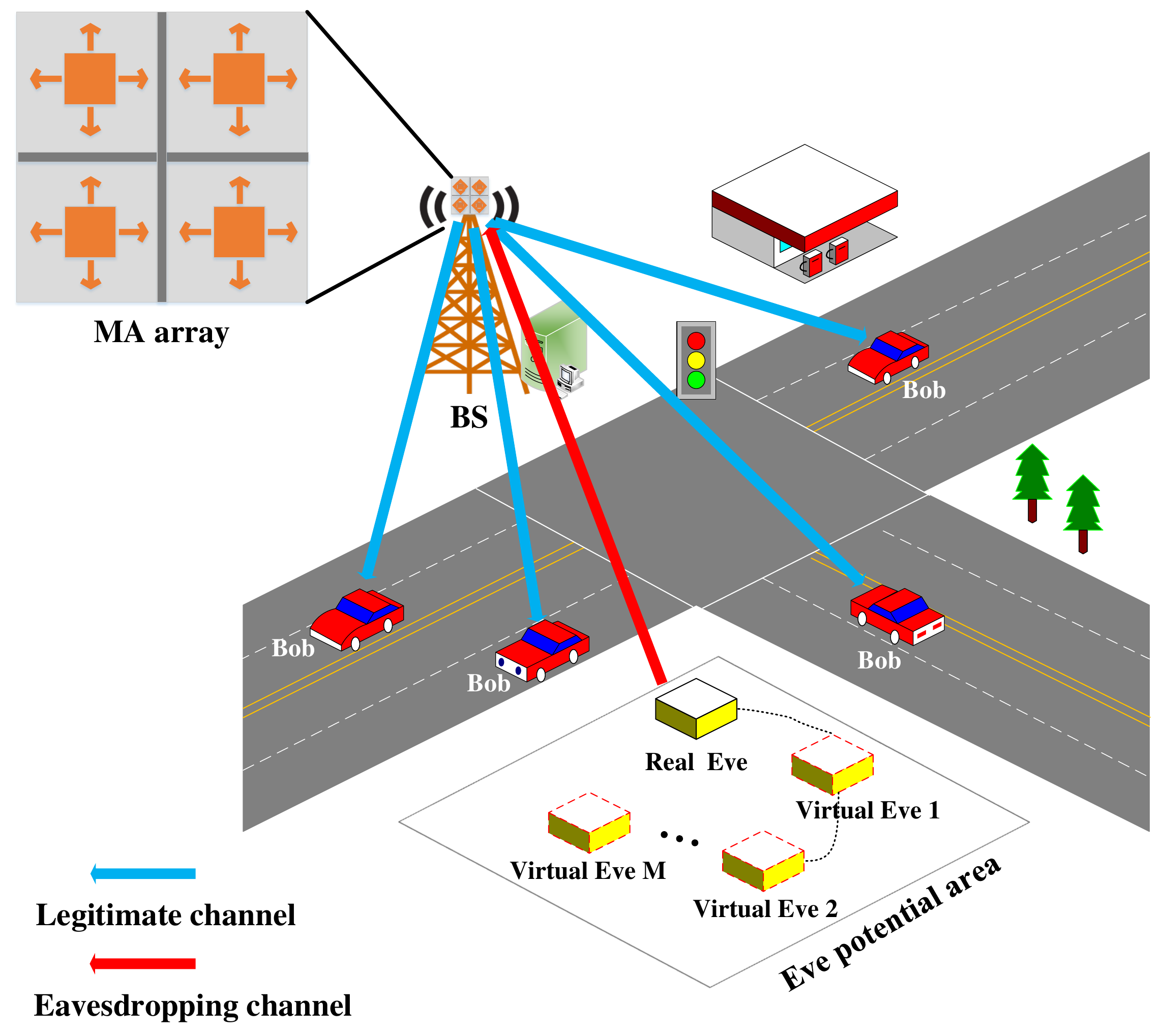}
  \caption{A MA-assisted secure communication system} 
  \label{fig:network model} 
\end{figure}

{We consider the downlink of a MA-assisted secure communication system that comprises on base station (BS), $K$ legitimate users (called as Bobs), and a potential Eve, as shown in Fig. \ref{fig:network model}. The BS is equipped with $N_t>1$ transmit MA, and sends the confidential information to \(K\) Bobs equipping with a single antenna. 
Different from the legitimate users, a single-antenna Eve may attempt deliberately eavesdrop the signals transmitted to the legitimate users via random movement. }

{The array of MA is at a height \(h\) from the ground, and MAs are connected to RF chains via flexible cables, and thus their positions can be adjusted in real time \cite{ref0.0011}. Without loss {of} generality, the $n$-th MA can move freely in a square region with the size of $A\times A$\footnote{The analyses can be extended to the case of three-dimensional moving space. In particular, the following analyses of the MA are based on three-dimensional moving space, and provide general guidances of enhancing the security of the system.}, and its position can be represented by Cartesian coordinates $\mathbf{t}_n = [x_n,y_n,z_n]^T$, under which $\Psi_n$ denotes the set of potential positions of {the} $n$-th MA due to the mobility.
The position matrix of $N$ MAs is represented as $\mathbf{T} = [\mathbf{t}_1, \mathbf{t}_2, \ldots, \mathbf{t}_N] \in \mathbb{R}^{3 \times N}$ for $n=1,...,N$. To avoid the coupling effect between any two MA, a minimum distance $d_{\min}$ is required as the secure distance between the antennas {\cite{ref0.0012}.} {Similarly, the position of the $k$-th Bob can be represented as $\mathbf{b}_k = [x_k,y_k,z_k]^T$, and the corresponding matrix is denoted as $\mathbf{B} = [\mathbf{b}_1, \mathbf{b}_2, \ldots, \mathbf{b}_K]$ for $k=1,...,K$.}}

\subsection{The CSI of the Bobs and Eve}
{Accurate estimation of CSI is crucial for the PLS design \cite{ref0.0013}. Accordingly, on the one hand, the Bobs can transmit pilot signals to the BS for facilitating channel estimation. As a result, the BS is able to periodically acquire the CSI of the Bobs. 
On the other hand, for the Eve, it sends signals to its dedicated wireless systems rather than the BS. Furthermore, as the Eve usually tries to hide its existence from the BS, it is not expected to cooperate with BS for CSI acquisition. Therefore, although the signal leakage from the Eve to the BS can still be utilized for channel estimation, {the acquired CSI is coarse and outdated \cite{ref0.0014}}. Consequently, some popular methods are adopted to characterize the CSI uncertainty (average CSI is known or perfect one is unknown), such as the deterministic model {\cite{ref0.0015,ref0.0016}}, statistical CSI error model \cite{ref0.0017}. However, due to the mobility of vehicles and random characteristics of wireless channels, the effectiveness and correctness of above CSI measuring models cannot be ensured.}

\subsubsection{MA-Assisted CSI Measurement Framework of Eve}\label{subsub:MA CSI Eve}
{To maximize the channel quality for intercepting the confidential information, the Eve may move to the eavesdropping location. In other words, the Eve moves rapidly within the potential area to maximize the interception benefit, and thus obtains the strongest strength of {the} eavesdropping signal, which is the worst-case scenario for ensuring the security of the system. In addition, the authors of \cite{ref0.0011} suggested that the moving vehicle within the potential area {could} be regarded as an available method to model the MA. Without loss of generality, the MA not only can enhance the channel quality between the BS and Bob, but also {can be} utilized to model the potential position of Eve. Next, we propose a CSI measurement framework of Eve based on the virtual MA as follows.}

\begin{figure}[htbp]
  \centering 
  \includegraphics[width=0.38\textwidth]{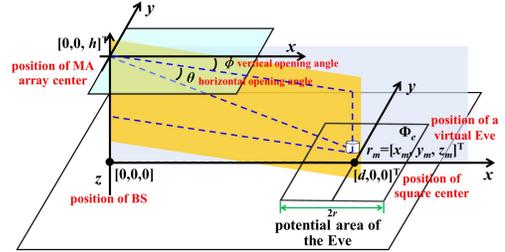}
  \caption{potential region model of Eve} 
  \label{fig:vitual MA of Eve}
\end{figure}

{Considering the CSI uncertainty of the Eve, for the sake of generality, its potential location area is modeled as a square region with the length of $2r$, as shown in Fig. \ref{fig:vitual MA of Eve}. Without loss of generality, we assume that $M$ potential positions of the Eve are constructed. In this way, the position of {the} $m$-th potential position of the Eve can be represented as $\mathbf{r}_m = \begin{bmatrix} x_m,y_m,z_m \end{bmatrix}^T$, and the position set of $M$ virtual Eves is denoted as \(\Phi_e\). To model all potential positions of the Eve, we can describe them by leveraging the mobility of the MA, and the Eve can rapidly move to the optimal eavesdropping position for intercepting the transmitted message.}

{Let $d$ be the distance between the square region's center and the BS, $h$ {represent} the height of the MA array from the ground, and the whole system be in a three-dimensional Cartesian coordinate. Without loss of generality, the position of the BS is set to the origin of $[0,0,0]$, then the coordinate of the square region's center is $[d, 0, 0]$. During the beamforming matrix design of the BS, the azimuth of the Eve relative to the BS affects the beamforming direction.
Typically, the horizontal opening angle and the vertical opening angle are used to describe the Eve's azimuth,  denoted by \( \theta \) and \( \phi \), respectively. Mathematically, the variation ranges of \( \theta \) and $\phi$ can be given by 
\begin{displaymath}
\theta\in\left[ \arctan\left(\frac{h}{d+r}\right), \arctan\left(\frac{h}{d-r}\right) \right]
\end{displaymath} 
and 
\begin{displaymath}
\phi\in\left[ -\arcsin\left(\frac{r}{f(d,r)}\right), \arcsin\left(\frac{r}{f(d,r)}\right) \right]
\end{displaymath} 
respectively, where $f(d,r)=\sqrt{(d-r)^2+ r^2}$ is a function of parameters $d$ and $r$. Accordingly, the general position expression of the Eve is given by $[\frac{h \cos \phi}{\tan \theta},\frac{h \sin \phi}{\tan \theta},0]^T$. That is, the position of the Eve is constrained by imposing restrictions on the angles, which is beneficial to the secure beamforming matrix design of the BS and position optimization of MAs.}

\subsection{Channel Model}
{Compared with the speed of wave propagation, the movement of the vehicles (i.e., Bobs and Eve) can be relatively ignored. In other words, the receivers {remain} relative stationary at the moment of downlink communication with the BS. Conversely, the channel between MA and receivers varies rapidly due to the speed of light scale and the moving distance of wavelength scale. For {the above reasons}, compared to the channel coherence time, the time that MAs take for the mobility can be also tolerable \cite{ref0.0018}. In this way, the channel matrix can be constant during a information transmission. 
It is noteworthy that for MA-enabled secure communications, the channel is re-configurable by adjusting the positions of MAs. Then, the BS-to-$k$-th Bob and the BS-to-$m$-th Eve channel vectors are given by $\mathbf{h}_k^b(\mathbf{T}) \in \mathbb{C}^{N \times 1} $ and $\mathbf{h}^e(\mathbf{T}, \mathbf{r}_m) \in \mathbb{C}^{N \times 1}$, respectively. Consequently, the expressions of received signals at the $k$-th Bob and {the} $m$-th virtual Eve can be expressed as 
\begin{equation}
y_k^b = \mathbf{h}_k^b(\mathbf{T})^H \left( \sum_{k=1}^K \mathbf{w}_k x_k \right) + n_k, \quad 1 \leq k \leq K
\end{equation}
and
\begin{equation}
y_m^e = \mathbf{h}^e(\mathbf{T}, \mathbf{r}_m)^H \left( \sum_{k=1}^K \mathbf{w}_k x_k \right) + n_e, \quad r_m \in \Phi_e
\end{equation}
respectively, where $(\cdot)^H$ is the conjugate transpose operation, the term of $\sum\limits_{k=1}^K \mathbf{w}_k x_k$ represents the cumulative signals sent by the BS to all $K$ Bobs, \(x_k\) is the symbol transmitted to the $k$-th Bob, \(\mathbf{w}_k \in \mathbb{C}^{N \times 1}\) is the corresponding beamforming vector, and the set of all beamforming vectors for all $K$ Bob is denoted as \(\mathbf{W} = [\mathbf{w}_1, \mathbf{w}_2, \ldots, \mathbf{w}_K] \in \mathbb{C}^{N \times K}\). The term of \(n_k\) is the thermal noise at the $k$-th Bob with $\sigma^2$ {being the} noise power. In particular, it follows a complex normal distribution with {zero mean} and variance $\sigma^2$.
Similarly, \(n_e\) is the thermal noise at Eve and \(n_e \sim \mathcal{CN}(0, \sigma^2)\).}

\begin{figure}[htbp]
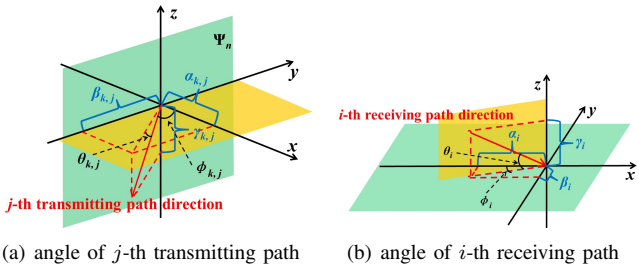

\centering  
\subfigure[angle of $j$-th transmitting path]{
\includegraphics[width=0.23\textwidth]{t3.pdf}\label{fig:spatial angle of j transmission path}}
\subfigure[angle of $i$-th receiving path]{
\includegraphics[width=0.23\textwidth]{t4.pdf}\label{fig:spatial angles different path Eve}}
\caption{Spatial angles of transmitting/receiving path}
\label{fig:transmit and receive path j i} 
\end{figure}

{In a MA-assisted communication system, the channel vector not only depends on the propagation environment, but also is related to the position of the MA. As a result, we consider the field-response based channel model, which assumes that the size of the transmit array is much smaller than the signal propagation distance, thereby satisfying the far-field assumption \cite{ref19,ref20}. 
Therefore, the MAs at different positions have the same angle of arrival (AoA), angle of departure (AoD), and complex amplitude coefficients{, where} only the phase of different channel paths changes with the MAs' positions \cite{ref21}, \cite{ref22}.
{To this end}, we first establish the field-response channel model of legitimate channel between the BS and the Bobs.
Given the $k$-th Bob (\( 1 \leq k \leq K \)), \( L_{k}^t \) {is} the number of all transmission paths between the MAs and it. We utilize the vertical angle \( \theta \) and horizontal angle \( \phi \) of each transmission path to characterize the direction of the $j$-th path (\( 1 \leq j \leq L_{k}^t \).). In detail, let \( \theta_{k,j} \) and \( \phi_{k,j} \) be the vertical and horizontal AoDs of the $j$-th transmission path at the MAs for the $k$-th Bob, respectively. In order to depict the projection of the normalized direction vector, we define the virtual AoD of {the} $j$-th transmission path in the three-dimensional coordinate system as the projection of the normalized direction vector on the three-dimensional coordinates, as shown in Fig. \ref{fig:spatial angle of j transmission path}.}

{Finally, the normalized three-dimensional spatial coordinates of the $j$-th transmission path, denoted by $\mathbf{p}_k^j=[\alpha_{k,j}, \beta_{k,j}, \gamma_{k,j}]^T$, can be derived as follows by using the trigonometric function, $\alpha_{k,j} = \cos \theta_{k,j} \cos \phi_{k,j}$, $\beta_{k,j}  = \cos \theta_{k,j} \sin \phi_{k,j}$, and $\gamma_{k,j} = \sin \theta_{k,j}$,
where \(\theta_{k,j} \in \left[-\frac{\pi}{2}, \frac{\pi}{2}\right] \) and \(\phi_{k,j} \in \left[-\frac{\pi}{2}, \frac{\pi}{2}\right]\) since each MA moves in the two-dimensional plane. Moreover, for the sake of convenience, the normalized direction vector of the $j$-th path in three-dimensional plane is projected into the Y-axis and Z-axis planes. In this way, when the $n$-th antenna moves to a new position, its position vector can be represented as \(\mathbf{t}_n = [0, y_n, z_n]^T\), and the distance difference between the position of the MA after movement and the original position for the $j$-th path is \(\mathbf{t}_n^T \mathbf{p}_k^j\), which can be calculated based on the fact that, given two vectors $\textbf{a}$ and $\textbf{b}$, the inner product of vectors $\textbf{a}$ and $\textbf{b}$ equals the product of the projection of vector $\textbf{a}$ onto vector $\textbf{b}$ and the magnitude of $\textbf{b}$, with the magnitude of the normalized vector $\textbf{b}$ being 1. In addition, the phase difference betweem two positions of the $j$-th path can be written as \(\frac{2\pi}{\lambda} \mathbf{t}_n^T \mathbf{p}_k^j\), where \(\lambda\) denotes the wavelength.
Therefore, given position coordinates of the MA, the corresponding transmit field response vector from the $n$-th MA to the $k$-th Bob is given by
\begin{equation}
g_{n,k}(\mathbf{t}_n) = \left[ e^{j\frac{2\pi}{\lambda} \mathbf{t}_n^T \mathbf{p}_k^1}, e^{j\frac{2\pi}{\lambda} \mathbf{t}_n^T \mathbf{p}_k^2}, \ldots, e^{j\frac{2\pi}{\lambda} \mathbf{t}_n^T \mathbf{p}_k^{L_k^t}} \right]^T \in \mathbb{C}^{{L_k^t} \times 1}
\end{equation}
and the transmit field response matrix from all MAs to {the} $k$-th user can be represented as
\begin{equation}
\mathbf{G}_k(\mathbf{T}) = [g_{1,k}(\mathbf{t}_1), g_{2,k}(\mathbf{t}_2), \ldots, g_{N,k}(\mathbf{t}_N)] \in \mathbb{C}^{{L_k^t} \times N} 
\end{equation}
While the Bob adopts a fixed antenna for the signal reception, the receive field response vector for the $k$-th Bob is given by 
\begin{equation}
\mathbf{f}_k = [1,1,\ldots,1]^T \in \mathbb{R}^{{L_k^t} \times 1}
\end{equation}
To better understand the relationship between the transmission path and the reception path, the path response matrix between all transmission paths of the MAs and {the} $k$-th Bob is defined as $\mathbf{\Sigma}_k = \text{diag}(\sigma_{k,1}, \sigma_{k,2}, ..., \sigma_{k,{L_k^t}}) \in \mathbb{C}^{{L_k^t} \times {L_k^t}}$, where $\sigma_{k,{L_k^t}}$ represents the complex amplitude response for a certain path. Thus, according to the channel response model, the field response channel matrix from the BS to the $k$-th Bob is 
\begin{equation}\label{eq:field response channel matrix bob}
\begin{aligned}
\mathbf{h}_k^{b}(\mathbf{T}) 
&= ((\mathbf{f}_k)^T\cdot\mathbf{\Sigma}_k\cdot\mathbf{G}_k(\mathbf{T}))^T = (\mathbf{G}_k(\mathbf{T}))^T\cdot(\mathbf{\Sigma}_k)^T\cdot\mathbf{f}_k \\
&= \left[ h_{k}^{b}(\mathbf{t}_1), h_{k}^{b}(\mathbf{t}_2), \ldots, h_{k}^{b}(\mathbf{t}_N) \right]^T \in \mathbb{C}^{N \times 1} 
\end{aligned}
\end{equation}
In particular, based on the conclusion of \cite{ref23}, the field response vector for the mobile antennas at different positions can be represented as the weighted sum of the responses of all paths, namely $h_k^{b}(\mathbf{t}) = \sum\limits_{l=1}^{L_k^t} \sigma_{k,l} e^{j \frac{2\pi}{\lambda} \mathbf{t}^T  \mathbf{p}_k^l}$.
}

{To sum up, the received Signal-to-Interference-plus-Noise Ratio (SINR) at the $k$-th Bob can be expressed as
\begin{equation}\label{eq:SINR bob}
\gamma_k^{b} = \frac{|\mathbf{h}_k^{b}(\mathbf{T})^H \mathbf{w}_k|^2}{\sum\limits_{k' \neq k} |\mathbf{h}_{k}^{b}(\mathbf{T})^H \mathbf{w}_{k'}|^2 + \sigma^2}
\end{equation}}

{As described in the Section \ref{subsub:MA CSI Eve}, the potential area including an Eve is modeled as a square region, and we propose a CSI measurement framework of Eve based on the virtual MA, which simulates $M$ potential positions of an Eve within the potential location area. Similar to Eq. \eqref{eq:field response channel matrix bob}, the field response matrix of the Eve transmitted by the BS is given by
\begin{equation}
\mathbf{G}(\mathbf{T}) = [g_{1}(\mathbf{t}_1), g_{2}(\mathbf{t}_2), \ldots, g_{N}(\mathbf{t}_N)] \in \mathbb{C}^{{L^t} \times N}
\end{equation}
Let $\mathbf{r}_m = \left[ x_m, y_m, 0 \right]^T$ be one of potential locations for the Eve by using {the} virtual MA, and \( L^r \) be the number of reception paths for the Eve. In addition, to better characterize the phase difference of the $i$-th path, the normalized direction vector for the $i$-th path from the BS to the Eve is given by $\textbf{p}^i = \begin{bmatrix}\cos \theta_{i} \cos \phi_{i}, \cos\theta_{i} \sin \phi_{i}, \sin \theta_{i}\end{bmatrix}^T$, as shown in Fig. \ref{fig:spatial angles different path Eve}, where \(\theta_i\) and \(\phi_i\) represent the vertical and horizontal AoAs for the $i$-th path between the BS and the Eve with constraints of \(\theta_{i} \in [0, \pi] \) and \(\phi_{i} \in \left[-\frac{\pi}{2}, \frac{\pi}{2}\right]\). Therefore, the received field response vector for the Eve located at a potential position \( \mathbf{r}_m \) can be represented as
\begin{equation}
\mathbf{f}^e(\mathbf{r}_m) = \left[ e^{j\frac{2\pi}{\lambda} \mathbf{r}_m^T \mathbf{p}^1}, e^{j\frac{2\pi}{\lambda} \mathbf{r}_m^T \mathbf{p}^2}, \ldots, e^{j\frac{2\pi}{\lambda} \mathbf{r}_m^T \mathbf{p}^{L^r}} \right]^T \in \mathbb{C}^{L^r \times 1}
\end{equation}
Next, for the ease of representation, the numbers of transmitting and receiving paths are the same (i.e., $L^r = L^t$). Let $\mathbf{\Sigma}_{b,e} = \text{diag}(\sigma_1, \dots, \sigma_{L^r})\in \mathbb{C}^{{L^r} \times {L^t}}$ be the path response matrix between the BS and the Eve. The case that transmitting and receiving paths have different size can be analyzed by using the similar method. Then, the corresponding field response channel matrix for the eavesdropping channel is given by 
\begin{equation}\label{eq:channel eve}
\begin{aligned}
\mathbf{h}^e(\mathbf{T}, \mathbf{r}_m) &= \left(( \mathbf{f}^e(\mathbf{r}_m) \right)^H \cdot\mathbf{\Sigma}_{b,e} \cdot\mathbf{G}(\mathbf{T}))^T \in \mathbb{C}^{N \times 1}\\
&= \left[ h^e(\mathbf{t}_1, \mathbf{r}_m), h^e(\mathbf{t}_2, \mathbf{r}_m), \ldots, h^e(\mathbf{t}_N, \mathbf{r}_m) \right]^T
\end{aligned}
\end{equation}}

{To sum up, when the Eve intercepts the confidential information of {the} $k$-th Bob sent by the BS, the corresponding received SINR at the Eve with the potential position of $\textbf{r}_m$ can be expressed as
\begin{equation}
\gamma_{m,k}^e = \frac{|\mathbf{h}^e(\mathbf{T}, \mathbf{r}_m)^H \mathbf{w}_k|^2}{\sum\limits_{k' \neq k} |\mathbf{h}^e(\mathbf{T}, \mathbf{r}_m)^H \mathbf{w}_{k'}|^2 + \sigma^2}
\end{equation}}

\subsection{Metrics of Physical Layer Security: Secrecy Rate}
{According to the SINR expression in Eq. \eqref{eq:SINR bob}, the achievable rate (bits/s/Hz) of $k$-th Bob can be given by
\begin{displaymath}
R_k^b = \log_2 (1 + \frac{|\mathbf{h}_k^{b}(\mathbf{T})^H \mathbf{w}_k|^2}{\sum\limits_{k' \neq k} |\mathbf{h}_{k}^{b}(\mathbf{T})^H \mathbf{w}_{k'}|^2 + \sigma^2})
\end{displaymath}
Then, for the security provisioning, we make a worst-case assumption regarding the capabilities of the Eve's mobility (i.e., CSI uncertainty or $M$ potential locations). In other words, the Eve can choose the best position to intercept the transmitted confidential information. Therefore, the channel capacity (bits/s/Hz) between the BS and the Eve at a potential location $\textbf{r}_m$ for decoding the signal of $k$-th Bob is given by
\begin{displaymath}
R_{m,k}^e = \log_2 (1 + \frac{|\mathbf{h}^e(\mathbf{T}, \mathbf{r}_m)^H \mathbf{w}_k|^2}{\sum\limits_{k' \neq k} |\mathbf{h}^e(\mathbf{T}, \mathbf{r}_m)^H \mathbf{w}_{k'}|^2 + \sigma^2})
\end{displaymath}
Hence, the maximum achievable secrecy rate between the BS and the $k$-th Bob is given by
\begin{equation}
R_{s,k} = \left[ R_k^b - R_{m,k}^e \right]^{+}
\end{equation} 
where $[u]^{+} = \max\{u, 0\}$. In particular, $R_{s,k}=0$ means that the legitimate channel is not secure, and no information will be transmitted. }

\subsection{Optimization Problem Formulation}
{In this paper, we aim to establish a worst-case robust secrecy rate maximization problem formulation for a CSI measurement framework of the Eve based on virtual MA (i.e., $M$ potential locations of an Eve). First, given any one of Bobs, denoted by {the} $k$-th Bob, and the Eve with the optimum location that can obtain the strongest signal of {the} $k$-th Bob, the corresponding secrecy rate is
\begin{displaymath}
R_{s,k} = \left[ \log_2 (1 + \gamma_k^{b}) - \max_{m \in \Phi_e} \left( \log_2 (1 + \gamma_{m,k}^e) \right) \right]^+
\end{displaymath}
To ensure the security of the whole system, we need  find the worst Bob with the lowest secrecy rate, and further maximize the corresponding secrecy rate. In other words, our goal is to maximize the minimum secrecy rate of the worst Bob for all potential positions of an Eve. That is,
\begin{displaymath}
\max \min_{k \in \Phi_b} R_{s,k}
\end{displaymath}
The joint design of the transmit beamforming at the BS and locations of the MA can be formulated as follows.
\begin{subequations}\label{eq:optimal inital}
\begin{align}
\text{Objective:}~~&\max_{\mathbf{T}, \mathbf{w}} \min_{k \in \Phi_b} R_{s,k} \label{eq:objective initial}\\
\text{s.t. } & \mathbf{t}_n \in \Psi_n \label{theparentequation b}\\
& \left| \mathbf{t}_n - \mathbf{t}_{n'} \right| \geq d_{\min}\label{theparentequation c}\\
& \theta \in \left[ \arctan\left(\frac{h}{d+r}\right), \arctan\left(\frac{h}{d-r}\right) \right] \label{theparentequation d}\\
& \phi \in \left[ -\arcsin\left(\frac{r}{f(d,r)}\right), \arcsin\left(\frac{r}{f(d,r)}\right) \right]\label{theparentequation e}\\
& \text{tr}(\mathbf{w}\mathbf{w}^H) \leq P_{\max} \label{theparentequation f}
\end{align}
\end{subequations}
In constraint \eqref{theparentequation f}, $P_{\max}$ is a non-negative parameter denoting the maximum transmit power of the BS. Constraint \eqref{theparentequation b} limits the moving region of the transmit MA, and constraint \eqref{theparentequation c} is the minimum moving distance difference between any two antennas in the MA array. Constraints \eqref{theparentequation d} and \eqref{theparentequation e} describe the potential moving area of the Eve with respect to two kinds of angles from the BS to it. Note that Eq. \eqref{eq:optimal inital} is non-convex optimization problem, due to the following reasons: 1) the 
operator $[\cdot]^+$ is non-smoothness for the objective function; 2) the non-convexity of objective function with respect to $(\textbf{T},\textbf{w})$ without the operation of $[\cdot]^+$; 3) the non-convex minimum distance constraint of \eqref{theparentequation c}. Moreover, the beamformer $\textbf{w}$ is coupled with $\textbf{T}$, which makes the problem more challenging to be solved.
To the best of the {authors'} knowledge, this is the first work that takes into account CSI uncertainty for the physical layer security enabled by the MA. Although there is no general approach to solving problem \eqref{eq:optimal inital} optimally, in the following section, we propose an effective algorithm to find a globally optimal solution} for problem \eqref{eq:optimal inital}.

\subsection{Optimization Problem Simplification}
{Through setting $\textbf{w}_k=0$, we can guarantee each item in $R_{s,k}$ be non-negative. Thus, the operation $[\cdot]^+$ can be securely omitted, and the optimization problem is transmitted into} 
\begin{subequations}\label{eq:optimal 2}
\renewcommand{\theequation}{\theparentequation\alph{equation}}
\begin{align}
\text{Objective:} ~~& \max_{\mathbf{T}, \mathbf{w}} \min_{k \in \Phi_b} \left[ \log_2 (1 + \gamma_k^b) - \max_{m \in \Phi_e} \log_2 (1 + \gamma_{m, k}^e) \right]\label{eq:2a} \\
\text{s.t.} ~~& \text{Eq. \eqref{theparentequation b}-Eq. \eqref{theparentequation f}}
\end{align}
\end{subequations}
In particular, problems \eqref{eq:optimal inital} and \eqref{eq:optimal 2} are equivalent \cite{ref24}.

{Although the objective in Eq. \eqref{eq:objective initial} is more tractable, it is difficult to solve since it is constructed by nested functions (i.e., maximum-minimum function), which cannot be simplified by using numerical methods \cite{ref25} or using first-order Taylor \cite{ref26}. Moreover, the optimization variable $ \mathbf{t}_n$ (namely the location of the $n$-th MA) is nested within the non-convex function $\mathbf{h}_k^{b}(\mathbf{T})$ and $\mathbf{h}^e(\mathbf{T}, \mathbf{r}_m) $, making it challenging to simplify the objective function through slack variables or numerical methods to solve this optimization problem. Therefore, different from current popular methods, the objective function can be simplified via exhaustive search. Then, due to the limited scale of sets \( \Phi_b \) and \( \Phi_e \), after specifying beamforming and MA's positions, we can find the Bob with the worst channel capacity and the Eve with the optimal eavesdropping position via exhaustive search, denoted by {the} \( \widetilde{k} \)-th Bob and {the} \( \widetilde{m} \)-th Eve, respectively. Thus, the optimization problem can be written by Eq. \eqref{eq:optimal 3}, as shown on the top of {the} next page.
\begin{figure*}
{\begin{subequations}\label{eq:optimal 3}
\renewcommand{\theequation}{\theparentequation\alph{equation}}
\begin{align}
\text{Objective:} ~~& \max_{\mathbf{T}, \mathbf{w}_{ \widetilde{k}}} \left( \log_2 \left( 1 + \frac{\left| \mathbf{h}_{ \widetilde{k}}^b(\mathbf{T})^H \mathbf{w}_{ \widetilde{k}} \right|^2}{\sum\limits_{k' \neq  \widetilde{k}} \left| \mathbf{h}_{ \widetilde{k}}^b(\mathbf{T})^H \mathbf{w}_{k'} \right|^2 + \sigma^2} \right) - \log_2 \left( 1 + \frac{\left| \mathbf{h}^e(\mathbf{T}, \mathbf{r}_{ \widetilde{m}})^H \mathbf{w}_{ \widetilde{k}} \right|^2}{\sum\limits_{k' \neq  \widetilde{k}} \left| \mathbf{h}^e(\mathbf{T}, \mathbf{r}_{ \widetilde{m}})^H \mathbf{w}_{k'} \right|^2 + \sigma^2} \right) \right)\label{eq:3a} \\
\text{s.t.} ~~& \text{Eq. \eqref{theparentequation b}, Eq. \eqref{theparentequation c}, Eq.\eqref{theparentequation d}, Eq. \eqref{theparentequation e}, and Eq. \eqref{theparentequation f}}
\end{align}
\end{subequations}}
\hrule 
\end{figure*}
}

\section{Optimization Algorithm Design for Secure MA-Assisted Wireless Communication}\label{sec:strategies}
{In this section, we foucs on solving the formulated optimization problem. Due to the tight coupling of $\textbf{w}$ and $\textbf{T}$, we tackle it via alternating optimization (AO). In fact, AO is a widely applicable and empirically efficient approach for handling optimization problems involving coupled optimization variables. It has been successfully applied to several secure wireless communication design problems, such as maximizing the achievable secrecy rate by joint design of beamforming and MA positions \cite{ref0.009}, and the average secrecy rate maximium by optimizing the trajectory of UAV, beamforming and phase shift of IRS \cite{ref27}. For the problem at hand, based on the principle of AO, we alternately solve for $(\textbf{w}, \textbf{T})$ while fixing the other variables. This yields a stationary point solution of problem \eqref{eq:optimal 3} as will be detailed in the following two subsections.}

\subsection{Transmit Beamformer Optimization for {A Fixed} $\textbf{T}$}

{First, we explore the design of $\textbf{w}$ by fixing $\textbf{T}$,
the subproblem of optimizing $\textbf{w}$ is given as follows.}

{\begin{subequations}
\renewcommand{\theequation}{\theparentequation\alph{equation}}
\begin{align}
\text{Objective:} ~~& \max_{\mathbf{w}_{ \widetilde{k}}} \Gamma(\mathbf{w}_{ \widetilde{k}})\label{eq:1a} \\
\text{s.t.} ~~& \text{Eq.\eqref{theparentequation d}, Eq. \eqref{theparentequation e}, and Eq. \eqref{theparentequation f}}
\end{align}
\end{subequations}
where $\Gamma(\mathbf{w}_ {\widetilde{k}}) = M(\mathbf{w}_{ \widetilde{k}}) - N(\mathbf{w}_{ \widetilde{k}})$, and 
\begin{displaymath}
 M(\mathbf{w}_{ \widetilde{k}}) =  \log_2 \left( 1 + \frac{\left| \mathbf{h}_{ \widetilde{k}}^b(\mathbf{T})^H \mathbf{w}_{ \widetilde{k}} \right|^2}{\sum\limits_{k' \neq  \widetilde{k}} \left| \mathbf{h}_{ \widetilde{k}}^b(\mathbf{T})^H \mathbf{w}_{k'} \right|^2 + \sigma^2} \right)
\end{displaymath}
\begin{displaymath}
N(\mathbf{w}_{ \widetilde{k}}) = \log_2 \left( 1 + \frac{\left| \mathbf{h}^e(\mathbf{T}, \mathbf{r}_{ \widetilde{m}})^H \mathbf{w}_{ \widetilde{k}} \right|^2}{\sum\limits_{k' \neq  \widetilde{k}} \left| \mathbf{h}^e(\mathbf{T}, \mathbf{r}_{ \widetilde{m}})^H \mathbf{w}_{k'} \right|^2 + \sigma^2} \right)
\end{displaymath}
In particular, $M(\mathbf{w}_{ \widetilde{k}})$ and $N(\mathbf{w}_{ \widetilde{k}})$ are convex functions with respect to $\mathbf{w}_{ \widetilde{k}}$, and the convexity of $\Gamma(\mathbf{w}_{ \widetilde{k}})$ is difficult to determine, since it is the difference of two convex functions. A projected gradient ascent (PGA) technique is then leveraged to tackle this non-convex problem. Then, the initial step involves determining the gradient of the objective function concerning the beamformer of the Bob with the worst channel capacity is given by Eq. \eqref{eq:initial gradient BF}, as shown on the top of {the} next page.
\begin{figure*}[t] 
{\begin{equation}\label{eq:initial gradient BF}
\nabla_{\mathbf{w}_{\widetilde{k}}} \Gamma(\mathbf{w}_{\widetilde{k}}) = \frac{1}{\ln 2} \left( 
\frac{\nabla_{\mathbf{w}_{\widetilde{k}}} |\mathbf{h}_{\widetilde{k}}^{b}(\mathbf{T})^H \mathbf{w}_{\widetilde{k}}|^2}{\sum\limits_{k' \neq \widetilde{k}} |\mathbf{h}_{\widetilde{k}}^{b}(\mathbf{T})^H \mathbf{w}_{k'}|^2 + \sigma^2+|\mathbf{h}_{\widetilde{k}}^{b}(\mathbf{T})^H \mathbf{w}_{\widetilde{k}}|^2}-  \frac{\nabla_{\mathbf{w}_{\widetilde{k}}} |\mathbf{h}^e(\mathbf{T}, \mathbf{r}_{\widetilde{m}})^H \mathbf{w}_{\widetilde{k}}|^2}{\sum\limits_{k' \neq \widetilde{k}}|\mathbf{h}^e(\mathbf{T}, \mathbf{r}_{\widetilde{m}})^H \mathbf{w}_{k'}|^2+\sigma^2+|\mathbf{h}^e(\mathbf{T}, \mathbf{r}_{\widetilde{m}})^H \mathbf{w}_{k'}|^2}  \right)
\end{equation}}
\hrule 
\end{figure*}
The composite gradient problems in Eq. \eqref{eq:initial gradient BF} are given by
\begin{displaymath}
\nabla_{\mathbf{w}_{ \widetilde{k}}} \left| \mathbf{h}_{ \widetilde{k}}^{b}(\mathbf{T})^H \mathbf{w}_{ \widetilde{k}} \right|^2 = 2\mathbf{h}_{ \widetilde{k}}^{b}(\mathbf{T}) \cdot \mathbf{h}_{ \widetilde{k}}^{b}(\mathbf{T})^H \mathbf{w}_{ \widetilde{k}}
\end{displaymath}
and
\begin{displaymath}
\nabla_{\mathbf{w}_{ \widetilde{k}}} \left| \mathbf{h}^e(\mathbf{T}, \mathbf{r}_ {\widetilde{m}})^H \mathbf{w}_{ \widetilde{k}} \right|^2 = 2\mathbf{h}^e(\mathbf{T}, \mathbf{r}_ {\widetilde{m}}) \cdot \mathbf{h}^e(\mathbf{T}, \mathbf{r}_ {\widetilde{m}})^H \mathbf{w}_{ \widetilde{k}}
\end{displaymath}
respectively. After each iteration, the beamformer \( \mathbf{w} \) for the worst-case Bob needs to be updated, and the updated value must comply with the power constraint \eqref{theparentequation f}. The update rule of beamformer  {based on AdaGrad algorithm \cite{ref28}} for the worst-case Bob in each iteration is presented as follows: 
\begin{equation}
\mathbf{w}_{ \widetilde{k}}^{v+1} = \mathbf{w}_{ \widetilde{k}}^{v} + \delta_{\text{adaw}} \nabla_{\mathbf{w}_{ \widetilde{k}}} \Gamma(\mathbf{w}_{ \widetilde{k}})
\end{equation}
\begin{equation}\label{eq:projection rules w}
\mathbf{w}_{ \widetilde{k}}^{v+1} = \Pi \left\{ \mathbf{w}_{ \widetilde{k}}^{v+1}, \mathbf{w}_{ \widetilde{k}}'\right\}
\end{equation}
where $\mathbf{w}_{ \widetilde{k}}' ={\mathbf{w}_{ \widetilde{k}}}\sqrt{ \frac{P_\text{max}}{q}}$, the parameter of \( q \) represents the power of the worst-case user after updating the beamforming vector. When the updated value of $\mathbf{w}_{ \widetilde{k}}'$ does not meet the power constraint \eqref{theparentequation f}, that is, when \( q > P_\text{max} \), the updated beamforming vector is projected proportionally to ensure compliance with the power constraint after the projection. Therefore, when the power constraint is not satisfied, \( \mathbf{w}_{ \widetilde{k}}' \) is used in place of \( \mathbf{w}_{ \widetilde{k}}^{v+1} \) for the next iteration. The pseudo-code of optimizing $\textbf{w}_{ \widetilde{k}}$ via PGA is shown in Algorithm \ref{alg:PGA beamformer}.}

\begin{algorithm}[!htp]
    \caption{PGA for optimizing $\textbf{w}_{ \widetilde{k}}$}
    \label{alg:PGA beamformer}
    \renewcommand{\algorithmicrequire}{\textbf{Initialize:}}
    \renewcommand{\algorithmicensure}{\textbf{Output:}}
    \begin{algorithmic}[1]
        \REQUIRE the maximum iteration number $I_{\rm ter}$, the maximum Monte Carlo simulation number $M_{w}$, the step size $\delta_{\text{adaw}}$, beamformer matrix $\textbf{W}$, and the convergence threshold $\tau_w$ 
        \REPEAT
            \STATE Compute the gradient value $\nabla_{\mathbf{w}_{ \widetilde{k}}} \Gamma(\mathbf{w}_{ \widetilde{k}})$  by Eq. \eqref{eq:initial gradient BF}
            \REPEAT
            \STATE  $t=t+\nabla_{\mathbf{w}_{ \widetilde{k}}} \Gamma(\mathbf{w}_{ \widetilde{k}})$
            \UNTIL{the maximum Monte Carlo simulation number $M_{w}$ is reached}
              \STATE Calculate the average gradient value $G_{\text{ave}} = t/M_{w}$
            \STATE Update beamformer $\mathbf{w}_{ \widetilde{k}}^{v+1} = \mathbf{w}_{ \widetilde{k}}^v+G_{\text{ave}}*\delta_{\text{adaw}}$
           
             \IF {$ \text{tr}(\mathbf{w}_{ \widetilde{k}}^{v+1}*(\mathbf{w}_{ \widetilde{k}}^{v+1})^H)  > P_\text{max}$}
             \STATE   $\mathbf{w}_{ \widetilde{k}}^{v+1}={\mathbf{w}_{ \widetilde{k}}^{v}}\sqrt{ \frac{P_\text{max}}{q}}$
             \ENDIF

        
                \UNTIL{convergence (i.e., $\left| \mathbf{w}_{ \widetilde{k}}^{v+1} - \mathbf{w}_{ \widetilde{k}}^{v} \right| < \tau_w$) or the maximum iteration number $I_{\rm ter}$ is reached}
        \STATE update $\mathbf{w}_{ \widetilde{k}}$ by the gradient value $\textbf{w}^{v+1}_{ \widetilde{k}}$
    \end{algorithmic}
\end{algorithm}

\subsection{Antenna Position Optimization for A Fixed $\textbf{w}_{ \widetilde{k}}$}
{Given a fixed beamformer $\textbf{w}_{ \widetilde{k}}$ obtained by Algorithm \ref{alg:PGA beamformer},  the antenna position optimization problem is formulated as 
\begin{subequations}\label{eq:optimization position initial}
\renewcommand{\theequation}{\theparentequation\alph{equation}}
\begin{align}
\text{Objective:} ~~& \max_{\mathbf{T}} F(\mathbf{T})\\
\text{s.t. } & \text{Eq. \eqref{theparentequation b}-Eq. \eqref{theparentequation e}}
\end{align}
\end{subequations}
where $F (\mathbf{T}) = P(\mathbf{T}) - Q(\mathbf{T})$, and 
\begin{displaymath}
 P(\mathbf{T}) = \log_2 \left( 1 + \frac{\left| \mathbf{h}_{ \widetilde{k}}^b(\mathbf{T})^H \mathbf{w}_{ \widetilde{k}} \right|^2}{\sum\limits_{k' \neq  \widetilde{k}} \left| \mathbf{h}_{ \widetilde{k}}^b(\mathbf{T})^H \mathbf{w}_{k'} \right|^2 + \sigma^2} \right) \nonumber 
\end{displaymath}
\begin{displaymath}
Q(\mathbf{T}) = \log_2 \left( 1 + \frac{\left| \mathbf{h}^e(\mathbf{T}, \mathbf{r}_{ \widetilde{m}})^H \mathbf{w}_{ \widetilde{k}} \right|^2}{\sum\limits_{k' \neq  \widetilde{k}} \left| \mathbf{h}^e(\mathbf{T}, \mathbf{r}_{ \widetilde{m}})^H \mathbf{w}_{k'} \right|^2 + \sigma^2} \right)  \label{eq:1a}
\end{displaymath}}

{The antenna position optimization for a fixed $\textbf{w}_{ \widetilde{k}}$ becomes challenging to solve, due to the tight coupling of the coordinates of $N$ MA, i.e., $\left\{ \mathbf{t}_n \right\}_{n=1}^N$. 
Specifically, we first divide a consistent trajectory $\textbf{T}$ into $N$ discrete blocks $\left\{ \mathbf{t}_n \right\}_{n=1}^N$. Then, we solve $N$ subproblems of \eqref{eq:optimization position initial}, which respectively optimize one transmit MA position $\textbf{t}_n$, with all the other variables being fixed. 
In this way, the coupling relationship among the positions of $N$ MAs can be decoupled, and further the position of each antenna can be optimized in sequence.
In addition, optimizing the position of the MA one by one can obtain a (at least) locally optimal solution for Eq. \eqref{eq:optimization position initial} by iteratively solving the above $N$ subproblems in an alternate manner.
The optimization of $\textbf{t}_n$ for the fixed $\left\{ \mathbf{t}_{n'} \right\}_{n'=1,n'\neq n}^N$ can be represented as follows.}
\begin{subequations}\label{eq:optimization position 2}
\renewcommand{\theequation}{\theparentequation\alph{equation}}
\begin{align}
\text{Objective:} ~~& \max_{ \mathbf{t}_n \in \Psi_n } F(\mathbf{t}_n;\mathbf{t}_{n'})\\
\text{s.t. } & \text{Eq. \eqref{theparentequation b}-Eq. \eqref{theparentequation e}}
\end{align}
\end{subequations}

{In particular, the position \( \mathbf{t}_n \) is embedded in the channel matrix, while the latter is non-convex, resulting in the optimization problem being challenging to solve. To solve the challenge, the stationary point of subproblem
\eqref{eq:optimization position 2} can be found by utilizing the PGA method. To this end, the projected gradient value $F(\mathbf{t}_n;\mathbf{t}_{n'})$ is given in Eq. \eqref{eq:gradient position tn}, as shown on the top of {the} next page.}
\begin{figure*}
{
\begin{equation}\label{eq:gradient position tn}
\begin{aligned}
\nabla_{\mathbf{t}_n} F(\mathbf{t}_n;\mathbf{t}_{n'}) = \frac{1}{\ln 2} \left[ X(\mathbf{t}_n;\mathbf{t}_{n'}) \cdot \nabla_{\mathbf{t}_n} \left( \frac{\chi(\mathbf{t}_n;\mathbf{t}_{n'})}{\alpha(\mathbf{t}_n;\mathbf{t}_{n'})} \right)  - Y(\mathbf{t}_n;\mathbf{t}_{n'}) \cdot \nabla_{\mathbf{t}_n} \left( \frac{\gamma(\mathbf{t}_n;\mathbf{t}_{n'})}{\beta(\mathbf{t}_n;\mathbf{t}_{n'})} \right) \right]
\end{aligned}
\end{equation}
where $X(\mathbf{t}_n;\mathbf{t}_{n'}) = \frac{\sum\limits_{k' \neq  \widetilde{k}} \left| \mathbf{h}_{ \widetilde{k}}^{b}(\mathbf{t}_n;\mathbf{t}_{n'})^H \mathbf{w}_{k'} \right|^2 + \sigma^2}{\left| \mathbf{h}_{ \widetilde{k}}^{b}(\mathbf{t}_n;\mathbf{t}_{n'})^H \mathbf{w}_{ \widetilde{k}} \right|^2 + \sum\limits_{k' \neq  \widetilde{k}} \left| \mathbf{h}_{ \widetilde{k}}^{b}(\mathbf{t}_n;\mathbf{t}_{n'})^H \mathbf{w}_{k'} \right|^2 + \sigma^2}$, $Y(\mathbf{t}_n;\mathbf{t}_{n'}) = \frac{\sum\limits_{k' \neq  \widetilde{k}} \left| \mathbf{h}^e(\mathbf{t}_n;\mathbf{t}_{n'}, \mathbf{r}_{ \widetilde{m}})^H \mathbf{w}_{k'} \right|^2 + \sigma^2}{\left| \mathbf{h}^e(\mathbf{t}_n;\mathbf{t}_{n'}, \mathbf{r}_{ \widetilde{m}})^H \mathbf{w}_{ \widetilde{k}} \right|^2 + \sum\limits_{k' \neq  \widetilde{k}} \left| \mathbf{h}^e(\mathbf{t}_n;\mathbf{t}_{n'}, \mathbf{r}_{ \widetilde{m}})^H \mathbf{w}_{k'} \right|^2 + \sigma^2}$, $\chi(\mathbf{t}_n;\mathbf{t}_{n'}) = \left| \mathbf{h}_{ \widetilde{k}}^{b}(\mathbf{t}_n;\mathbf{t}_{n'})^H \mathbf{w}_{ \widetilde{k}} \right|^2$, $\gamma(\mathbf{t}_n;\mathbf{t}_{n'}) = \left| \mathbf{h}^e(\mathbf{t}_n;\mathbf{t}_{n'}, \mathbf{r}_{ \widetilde{m}})^H \mathbf{w}_{ \widetilde{k}} \right|^2$, $\alpha(\mathbf{t}_n;\mathbf{t}_{n'}) = \sum\limits_{k' \neq  \widetilde{k}} \left| \mathbf{h}_{ \widetilde{k}}^{b}(\mathbf{t}_n;\mathbf{t}_{n'})^H \mathbf{w}_{k'} \right|^2 + \sigma^2$, $\beta(\mathbf{t}_n;\mathbf{t}_{n'}) = \sum\limits_{k' \neq  \widetilde{k}} \left| \mathbf{h}^e(\mathbf{t}_n;\mathbf{t}_{n'}, \mathbf{r}_{ \widetilde{m}})^H \mathbf{w}_{k'} \right|^2 + \sigma^2$}
\hrule 
\end{figure*}
By using the chain rule, the composite gradients values of Eq. \eqref{eq:gradient position tn} are given by the top of next page,
\begin{figure*}[t] 
\begin{equation}\label{eq:two gradient chi alpha}
\nabla_{\mathbf{t}_n} \left( \frac{\chi(\mathbf{t}_n;\mathbf{t}_{n'})}{\alpha(\mathbf{t}_n;\mathbf{t}_{n'})} \right) = \frac{\nabla_{\mathbf{t}_n} \chi(\mathbf{t}_n;\mathbf{t}_{n'}) \cdot \alpha(\mathbf{t}_n;\mathbf{t}_{n'}) - \nabla_{\mathbf{t}_n} \alpha(\mathbf{t}_n;\mathbf{t}_{n'}) \cdot \chi(\mathbf{t}_n;\mathbf{t}_{n'})}{\alpha(\mathbf{t}_n;\mathbf{t}_{n'})^2}
\end{equation}
\begin{equation}\label{eq:two gardient gamma beta}
\nabla_{\mathbf{t}_n} \left( \frac{\gamma(\mathbf{t}_n;\mathbf{t}_{n'})}{\beta(\mathbf{t}_n;\mathbf{t}_{n'})} \right) = \frac{\nabla_{\mathbf{t}_n} \gamma(\mathbf{t}_n;\mathbf{t}_{n'}) \cdot \beta(\mathbf{t}_n;\mathbf{t}_{n'}) - \nabla_{\mathbf{t}_n} \beta(\mathbf{t}_n;\mathbf{t}_{n'}) \cdot \gamma(\mathbf{t}_n;\mathbf{t}_{n'})}{\beta(\mathbf{t}_n;\mathbf{t}_{n'})^2}
\end{equation}
\hrule 
\end{figure*}
respectively. Next, we calculate the gradient values of $\chi(\mathbf{t}_n;\mathbf{t}_{n'})$ and $\alpha(\mathbf{t}_n;\mathbf{t}_{n'})$ in Eq. \eqref{eq:two gradient chi alpha}. Let $\Psi_{ \widetilde{k}}(\mathbf{t}_n;\mathbf{t}_{n'})=\mathbf{h}_{ \widetilde{k}}^{b}(\mathbf{t}_n;\mathbf{t}_{n'})^H \mathbf{w}_{ \widetilde{k}}$. By using the chain rule, we get 
\begin{displaymath}
\nabla_{\mathbf{t}_n} \chi(\mathbf{t}_n;\mathbf{t}_{n'}) = \nabla_{\mathbf{t}_n} \left| \Psi_{ \widetilde{k}}(\mathbf{t}_n;\mathbf{t}_{n'}) \right|^2 = 2 \Psi_{ \widetilde{k}}(\mathbf{t}_n;\mathbf{t}_{n'})  \cdot \mathbf{J}_{h_b} \cdot \mathbf{w}_{ \widetilde{k}}
\end{displaymath}
where $\mathbf{J}_{h_b}$ denotes the Jacobian matrix of \( \mathbf{h}_{ \widetilde{k}}^{b}(\mathbf{t}_n;\mathbf{t}_{n'})^H \), and can be represented as
\begin{displaymath}
\begin{aligned}
\mathbf{J}_{h_b} 
&= \begin{bmatrix}
\frac{\partial \mathbf{h}_{ \widetilde{k}}^{b}(\mathbf{t}_n;\mathbf{t}_{n'})^H}{\partial x_n} & \frac{\partial \mathbf{h}_{ \widetilde{k}}^{b}(\mathbf{t}_n;\mathbf{t}_{n'})^H}{\partial y_n} & \frac{\partial \mathbf{h}_{ \widetilde{k}}^{b}(\mathbf{t}_n;\mathbf{t}_{n'})^H}{\partial z_n} \\
\end{bmatrix}^T\\
 &=
\begin{bmatrix}
\frac{\partial h_{ \widetilde{k}}^{b}(\mathbf{t}_1)^*}{\partial x_n} & \cdots &
\frac{\partial h_{ \widetilde{k}}^{b}(\mathbf{t}_n)^*}{\partial x_n} & \cdots & \frac{\partial h_{ \widetilde{k}}^{b}(\mathbf{t}_N)^*}{\partial x_n} \\
\frac{\partial h_{ \widetilde{k}}^{b}(\mathbf{t}_1)^*}{\partial y_n} & \cdots &
\frac{\partial h_{ \widetilde{k}}^{b}(\mathbf{t}_n)^*}{\partial y_n} & \cdots & \frac{\partial h_{ \widetilde{k}}^{b}(\mathbf{t}_N)^*}{\partial y_n} \\
\frac{\partial h_{ \widetilde{k}}^{b}(\mathbf{t}_1)^*}{\partial z_n} & \cdots &
\frac{\partial h_{ \widetilde{k}}^{b}(\mathbf{t}_n)^*}{\partial z_n} & \cdots & \frac{\partial h_{ \widetilde{k}}^{b}(\mathbf{t}_N)^*}{\partial z_n}
\end{bmatrix}
\end{aligned}
\end{displaymath}
where $*$ denotes the conjugate of $h_{ \widetilde{k}}^{b}(\mathbf{t})$. In fact, partial derivatives in the Jacobian matrix holds, if and only if the partial derivative of the  \( \mathbf{h}_{ \widetilde{k}}^{b}(\mathbf{t}_n;\mathbf{t}_{n'})^H \) with respect to the position of the $n$-th antenna exists, and the other elements of the corresponding Jacobian matrix are zero. Therefore, the partial derivatives with respect to the three coordinates of the $n$-th antenna are given by Eq. \eqref{eq:partial three coordinates}, as shown on the top of {the} next page.
\begin{figure*}
{
\begin{equation}\label{eq:partial three coordinates}
\begin{aligned}
\frac{\partial h_{ \widetilde{k}}^{b}(\mathbf{t}_n)^*}{\partial x_n} &=
\frac{2 \pi}{\lambda} \sum_{l=1}^{L_k^t} \sigma_{k,l}\cos\theta_{k,l} \cos\phi_{k,l} \left( -\sin\left(\frac{2\pi}{\lambda} \mathbf{t}_n^T \mathbf{p}_{k}^{l}\right) - j \cos\left(\frac{2\pi}{\lambda} \mathbf{t}_n^T \mathbf{p}_{k}^{l}\right) \right)\\
\frac{\partial h_{ \widetilde{k}}^{b}(\mathbf{t}_n)^*}{\partial y_n} &=
\frac{2 \pi}{\lambda} \sum_{l=1}^{L_k^t} \sigma_{k,l} \cos\theta_{k,l} \sin\phi_{k,l} \left( -\sin\left(\frac{2\pi}{\lambda} \mathbf{t}_n^T \mathbf{p}_{k}^{l}\right) - j \cos\left(\frac{2\pi}{\lambda} \mathbf{t}_n^T \mathbf{p}_{k}^{l}\right) \right)\\
\frac{\partial h_{ \widetilde{k}}^{b}(\mathbf{t}_n)^*}{\partial z_n} &= 
\frac{2 \pi}{\lambda} \sum_{l=1}^{L_k^t} \sigma_{k,l}  \sin\phi_{k,l}\left( -\sin\left(\frac{2\pi}{\lambda} \mathbf{t}_n^T \mathbf{p}_{k}^{l}\right) - j \cos\left(\frac{2\pi}{\lambda} \mathbf{t}_n^T \mathbf{p}_{k}^{l}\right) \right)
\end{aligned}
\end{equation}}
\hrule 
\end{figure*}
The gradient expression of \( \nabla_{\mathbf{t}_n} \chi(\mathbf{t}_n;\mathbf{t}_{n'}) \) can be derived. Let $\Omega_{k'}(\mathbf{t}_n;\mathbf{t}_{n'} ) = \mathbf{h}_{ \widetilde{k}}^{b}(\mathbf{t}_n;\mathbf{t}_{n'})^H \mathbf{w}_{k'}$. Similarly, the gradient \( \nabla_{\mathbf{t}_n} \alpha(\mathbf{t}_n;\mathbf{t}_{n'}) \) is given by
\begin{displaymath}
\nabla_{\mathbf{t}_n} \alpha(\mathbf{t}_n;\mathbf{t}_{n'}) 
= 2 \sum_{k' \neq  \widetilde{k}} \Omega_{k'}(\mathbf{t}_n;\mathbf{t}_{n'}) \cdot \mathbf{J}_{h_b} \cdot \mathbf{w}_{k'}
\end{displaymath}
{We take the real part of the above two  partial derivatives and obtain the gradient Eq. \eqref{eq:two gradient chi alpha}. }Next, we calculate the gradients $\nabla_{\mathbf{t}_n} \gamma(\mathbf{t}_n;\mathbf{t}_{n'})$ and $\nabla_{\mathbf{t}_n} \beta(\mathbf{t}_n;\mathbf{t}_{n'})$ in Eq. \eqref{eq:two gardient gamma beta}. According to the  result in Eq. \eqref{eq:channel eve}, we first get
\begin{displaymath}
h^e(\mathbf{t}, \mathbf{r}_m) = \sum_{u=1}^{L^r} \sigma_{u} e^{j \frac{2\pi}{\lambda} (\mathbf{t}^T\mathbf{p}_k^u - \mathbf{r}_m^T \mathbf{p}^u)}
\end{displaymath}
Then, the gradients \( {\gamma(\mathbf{t}_n;\mathbf{t}_{n'})} \) and \( {\beta(\mathbf{t}_n;\mathbf{t}_{n'})} \) can be derived by using the same method in Eq. \eqref{eq:two gardient gamma beta}, where  $\Theta_{ \widetilde{k}}(\mathbf{t}_n;\mathbf{t}_{n'}) = \mathbf{h}^e(\mathbf{t}_n;\mathbf{t}_{n'}, \mathbf{r}_{ \widetilde{m}})^H \mathbf{w}_{ \widetilde{k}}$, and $O_{k'}(\mathbf{t}_n;\mathbf{t}_{n'}) = \mathbf{h}^e(\mathbf{t}_n;\mathbf{t}_{n'}, \mathbf{r}_{ \widetilde{m}})^H \mathbf{w}_{k'}$.
\begin{equation}
\label{eq:gradient gamma}
\nabla_{\mathbf{t}_n} \gamma(\mathbf{t}_n;\mathbf{t}_{n'}) = 2   \Theta_{ \widetilde{k}}(\mathbf{t}_n;\mathbf{t}_{n'}) \cdot \mathbf{J}_{h_e} \cdot \mathbf{w}_{ \widetilde{k}} 
\end{equation}
\begin{equation}
\label{eq:gradient beta}
\nabla_{\mathbf{t}_n} \beta(\mathbf{t}_n;\mathbf{t}_{n'}) = \sum_{k' \neq  \widetilde{k}} 2  O_{k'}(\mathbf{t}_n;\mathbf{t}_{n'}) \cdot \mathbf{J}_{h_e} \cdot \mathbf{w}_{k'} 
\end{equation}
where \(\mathbf{J}_{h_e}\) denotes the Jacobian matrix of $\mathbf{h}^e(\mathbf{t}_n;\mathbf{t}_{n'}, \mathbf{r}_m)^H$ and can be represented as
\begin{equation}
\begin{aligned}
\mathbf{J}_{h_e} 
&= \begin{bmatrix}
\frac{\partial \mathbf{h}^e(\mathbf{t}_n;\mathbf{t}_{n'},  \mathbf{r}_{\widetilde{m}})^H}{\partial x_n} & \frac{\partial \mathbf{h}^e(\mathbf{t}_n;\mathbf{t}_{n'}, \mathbf{r}_{\widetilde{m}})^H}{\partial y_n} & \frac{\partial \mathbf{h}^e(\mathbf{t}_n;\mathbf{t}_{n'}, \mathbf{r}_{\widetilde{m}})^H}{\partial z_n}
\end{bmatrix}^T\\
&= \begin{bmatrix}
\frac{\partial h^e(\mathbf{t}_{1}, \mathbf{r}_{\widetilde{m}})^*}{\partial x_n} & \dots & \frac{\partial h^e(\mathbf{t}_{n}, \mathbf{r}_{\widetilde{m}})^*}{\partial x_n} & \dots & \frac{\partial h^e(\mathbf{t}_{N}, \mathbf{r}_{\widetilde{m}})^*}{\partial x_n} \\
\frac{\partial h^e(\mathbf{t}_{1}, \mathbf{r}_{\widetilde{m}})^*}{\partial y_n} & \dots &  \frac{\partial h^e(\mathbf{t}_{n}, \mathbf{r}_{\widetilde{m}})^*}{\partial y_n} & \dots & \frac{\partial h^e(\mathbf{t}_{N}, \mathbf{r}_{\widetilde{m}})^*}{\partial y_n} \\
\frac{\partial h^e(\mathbf{t}_{1}, \mathbf{r}_{\widetilde{m}})^*}{\partial z_n}& \dots &  \frac{\partial h^e(\mathbf{t}_{n}, \mathbf{r}_{\widetilde{m}})^*}{\partial z_n} & \dots & \frac{\partial h^e(\mathbf{t}_{N}, \mathbf{r}_{\widetilde{m}})^*}{\partial z_n}
\end{bmatrix}
\end{aligned}
\end{equation}
Consequently, the partial derivatives with respect to the three coordinates of the $n$-th antenna are given by Eq. \eqref{eq:partial three coordinates2}, as shown on the top of {the} next page.
\begin{figure*}
{
\begin{equation}\label{eq:partial three coordinates2}
\frac{\partial h^e(\mathbf{t}_{n}, \mathbf{r}_{\widetilde{m}})^*}{\partial x_n} = \frac{2\pi}{\lambda} \sum_{u=1}^{L^r} \sigma_{u} \cos\theta_{u} \cos\phi_{u} \left( -\sin\left(\frac{2\pi}{\lambda} (\mathbf{t}_n^T \mathbf{p}_{k}^{u} - \mathbf{r}_{\widetilde{m}}^T \mathbf{p}^{u})\right) - j \cos\left(\frac{2\pi}{\lambda} (\mathbf{t}_n^T \mathbf{p}_{k}^{u} - \mathbf{r}_{\widetilde{m}}^T \mathbf{p}^{u})\right) \right)
\end{equation}
\begin{equation}
\frac{\partial h^e(\mathbf{t}_{n}, \mathbf{r}_{\widetilde{m}})^*}{\partial y_n} = \frac{2\pi}{\lambda} \sum_{u=1}^{L^r} \sigma_{u} \cos\theta_{u} \sin\phi_{u} \left( -\sin\left(\frac{2\pi}{\lambda} (\mathbf{t}_n^T \mathbf{p}_{k}^{u} - \mathbf{r}_{\widetilde{m}}^T \mathbf{p}^{u})\right) - j \cos\left(\frac{2\pi}{\lambda} (\mathbf{t}_n^T \mathbf{p}_{k}^{u} - \mathbf{r}_{\widetilde{m}}^T \mathbf{p}^{u})\right) \right)
\end{equation}
\begin{equation}
\frac{\partial h^e(\mathbf{t}_{n}, \mathbf{r}_{\widetilde{m}})^*}{\partial z_n} = \frac{2\pi}{\lambda} \sum_{u=1}^{L^r} \sigma_{u} \sin\phi_{u} \left( -\sin\left(\frac{2\pi}{\lambda} (\mathbf{t}_n^T \mathbf{p}_{k}^{u} - \mathbf{r}_{\widetilde{m}}^T \mathbf{p}^{u})\right) - j \cos\left(\frac{2\pi}{\lambda} (\mathbf{t}_n^T \mathbf{p}_{k}^{u} - \mathbf{r}_{\widetilde{m}}^T \mathbf{p}^{u})\right) \right)
\end{equation}}
\hrule 
\end{figure*}
By employing the same method to take the real part of the gradients  Eq. \eqref{eq:gradient gamma} and Eq. \eqref{eq:gradient beta} and can obtain the gradient Eq. \eqref{eq:two gardient gamma beta}. Finally the gradient expression of $F(\mathbf{t}_n;\mathbf{t}_{n'})$ can be derived.

{When finding the optimization position of the $n$-th MA via the PGA method, {based on AdaGrad algorithm}, {the} corresponding position in $(v+1)$-th iteration {is firstly} given by
\begin{equation}\label{eq:position update initial}
(\mathbf{t}_n)^{v+1} = (\mathbf{t}_n)^v + \delta_{\rm adat} \cdot \nabla_{\mathbf{t}_n} F(\mathbf{t}_n;\mathbf{t}_{n'})
\end{equation}
Then, we need respectively determine whether the constraints in Eq. \eqref{theparentequation b} and Eq. \eqref{theparentequation c} are true or not. If the conditions hold, the value of $\textbf{t}_n$ in $(v+1)$-th iteration keeps the same. Following the idea of MA's position updating based on ULA \cite{ref0.009}, the current position update of the $n$-th MA only depends on its previous state. Based on the assumption, when the distance between the $n$-th MA after moving and the $(n-1)$-the MA does not satisfy the minimum distance constraint \eqref{theparentequation b}, the value of $\textbf{t}_n$ in $(v+1)$-th iteration can be updated as
\begin{equation}\label{eq:rule2}
(\mathbf{t}_n)^{v+1} = \mathbf{t}_{n-1} + d_{\text{min}} \cdot \frac{(\mathbf{t}_n)^{v+1} - \mathbf{t}_{n-1}}{\| (\mathbf{t}_n)^{v+1} - \mathbf{t}_{n-1} \|}
\end{equation}
In other words, when the position updated by Eq. \eqref{eq:position update initial} does not meet the minimum distance constraint \eqref{theparentequation c}, a projection method will be employed. Recall that $d_{\min}$ is the minimum distance between any two MAs, the 
the position of the $n$-th MA in {the} $(v+1)$-th iteration can be updated by projecting it to a circle with the radius of $d_{\min}$, along the direction from the position of the $(n-1)$-th MA pointing to 
the $n$-th MA updating in {the} $(v+1)$-th iteration, which can ensure that the minimum distance constraint Eq. \eqref{theparentequation c}.}
{While the position of the $n$-th MA in the $(v+1)$-th iteration exceeds the range of $\Psi_n$, a projection method will be employed.  
Let \( \left[ x_{\text{min}}, x_{\text{max}} \right] \times \left[ y_{\text{min}}, y_{\text{max}} \right] \times \left[ z_{\text{min}}, z_{\text{max}} \right]
 \) be the movement range for the \( n \)-th MA. Then the three-dimensional coordinates of the position $\textbf{t}_n$ in the $(v+1)$-th iteration are projected as follows: 
\begin{displaymath}
\begin{aligned}
x_n' &= \min(\max(x_n^{v+1}, x_{\text{min}}), x_{\text{max}})\\
y_n' &= \min(\max(y_n^{v+1}, y_{\text{min}}),y_{\text{max}}) \\
z_n' &= \min(\max(z_n^{v+1}, z_{\text{min}}),z_{\text{max}})  
\end{aligned}
\end{displaymath} 
Finally, when the antenna position $\textbf{t}_n$ after being updated based on Eq. \eqref{eq:position update initial} is out of range $\Psi_n$, it will be projected into red the closest boundary of a mobile area of {the} $n$-th MA, thus the final position of the $n$-th MA in the $(v+1)$-th iteration is
\begin{equation}\label{eq:rule3}
(\mathbf{t}_n)^{v+1} = [x_n', y_n', z_n']^T
\end{equation}
To sum up, the pseudo-code of optimizing $\textbf{t}_n$ via the PGA method is shown in Algorithm \ref{alg:PGA antenna position}.}

\begin{algorithm}[!htp]
    \caption{PGA for optimizing $\textbf{t}_{n}$}
    \label{alg:PGA antenna position}
    \renewcommand{\algorithmicrequire}{\textbf{Initialize:}}
    \renewcommand{\algorithmicensure}{\textbf{Output:}}
    \begin{algorithmic}[1]
        \REQUIRE the maximum iteration number $I_{\rm ter}$, the maximum Monte Carlo simulation number $M_t$, the step size $\delta_{\text{adat}}$, and the convergence threshold $\tau_T$ 
         \FOR{each $\mathbf{t}_n$}
            \REPEAT
                 \STATE Compute the gradient value $\nabla_{\mathbf{t}_{n}} F(\mathbf{t}_{n};\mathbf{t}_{n'})$  by Eq. \eqref{eq:initial gradient BF}
                 \REPEAT
                 \STATE  $t=t+\nabla_{\mathbf{t}_{n}} F(\mathbf{t}_{n};\mathbf{t}_{n'})$
                 \UNTIL{the maximum Monte Carlo simulation number $M_t$ is reached}
                 \STATE Calculate the average gradient value $G_{\text{ave}} = t/M_t$
            
                 \STATE Update  $\mathbf{t}_{n}^{v+1} = \mathbf{t}_{n}^v+G_{\text{ave}}*\delta_{\text{adat}}$
                \IF {$\mathbf{t}_{n}^{v+1}$ does not satisfied the s.t. \eqref{theparentequation b} }
                 \STATE Update  $\mathbf{t}_{n}^{v+1} = [x_n', y_n', z_n']^T$
                \ENDIF
                \IF {$\mathbf{t}_{n}^{v+1}$ does not {satisfy} the s.t. \eqref{theparentequation c} }
                 \STATE Update $\mathbf{t}_{n}^{v+1} = \mathbf{t}_{n-1} + d_{\text{min}} \cdot \frac{(\mathbf{t}_n)^{v+1} - \mathbf{t}_{n-1}}{\| (\mathbf{t}_n)^{v+1} - \mathbf{t}_{n-1} \|} $
                \ENDIF
                \UNTIL{convergence (i.e., $\left| \mathbf{t}_{n}^{v+1} - \mathbf{t}_{n}^{v} \right| < \tau_T$) or the maximum iteration number $I_{\rm ter}$ is reached}
              \ENDFOR  
        \STATE update $\mathbf{t}_{n}$ by final gradient value $\mathbf{t}_{n}^{v+1}$
    \end{algorithmic}
\end{algorithm}

{In above analyses of optimization problem \eqref{eq:optimal inital}, we have derived the optimal beamforming matrix of the BS and positions of all MAs, respectively. However, due to the non-convex nature of the optimization problem \eqref{eq:optimal inital}, solely using methods of PGA and AO may lead to the optimization process to get stuck in a local optimal solution or even does not reach the convergence. Therefore, in the following subsection, combining with the simulated annealing (SA), a heuristic search method, we can search more feasible solutions in a wider space by accepting inferior solutions with a {high} probability to avoid getting stuck in local optimal solutions.}

\subsection{{Proposed  SA-PGA Algorithm}}
{In this subsection, we propose a SA-PGA algorithm to solving {the} complex non-convex problem \eqref{eq:optimal inital}. The principle of the SA algorithm is to simulate the real annealing process of metals. When the iteration just begins, i.e., being the state of ``high temperatures'', the SA can accept solutions worse than the previous iteration with a certain probability. As the iteration proceeds, the ``high temperature'' of metals gradually decreases, which also means the algorithm’s probability of accepting worse solutions becomes lower, and the demand for the quality of solutions becomes higher, aiming to search for the optimal solution within a larger scope as much as possible, and finally reach a stable state. Furthermore, combining SA and PGA can effectively solve complexity of problem \eqref{eq:optimal inital} that requires a global search under strict constraints. The proposed SA-PGA method improves the possibility of finding a global optimum while ensuring the feasibility and efficiency of the solution. Finally, the pseudo-code of the SA-PGA is shown in Algorithm \ref{alg:SA-PGA}. The main process of SA-PGA is described as follows.}

\begin{algorithm}[!htp]
\label{alg:SA-PGA}
    \caption{{Overall algorithm of SA-PGA}}
    \label{alg:SA-PGA}
    \renewcommand{\algorithmicrequire}{\textbf{Initialize:}}
    \renewcommand{\algorithmicensure}{\textbf{Output:}}
    \begin{algorithmic}[1]
        \REQUIRE the maximum iteration number $I_{\rm ter}$, {initial temperature $T$, temperature decreasing coefficient $\beta$, probability $\rho$}, previous position of antenna $\mathbf{t}_{\text{previous}}$, previous beamforming vector $\mathbf{w}_{\text{previous}}$, previous secrecy rate $R_{\text{previous}}$
            \REPEAT
                \STATE $\mathbf{w}_{\text{op}} \gets$ outputted by Algorithm \ref{alg:PGA beamformer} with $\mathbf{w}_{\text{previous}}$
                \STATE $\mathbf{t}_{\text{op}} \gets$ outputted by Algorithm \ref{alg:PGA antenna position} with $\mathbf{t}_{\text{previous}}$                \STATE calculate the optimal secrecy rate $R_{\text{op}}$
            \IF{{$R_{\text{op}}>R_{\text{previous}}$ \text{or} accept $R_{\text{op}}$ with a probability of $\rho$}}
                \STATE $\mathbf{t}_{\text{previous}} \gets \mathbf{t}_{\text{op}}$, $\mathbf{w}_{\text{previous}} \gets \mathbf{w}_{\text{op}}$, $R_{\text{previous}} \gets R_{\text{op}}$
                \ENDIF
                \STATE $T = T \cdot \beta$
                \UNTIL{ the maximum iteration number $I_{\rm ter}$ is reached}  
       \ENSURE $\mathbf{t}_{\text{previous}}, \mathbf{w}_{\text{previous}}$   
    \end{algorithmic}
\end{algorithm}

{The iterative process begins by applying PGA to find potential optimal values for the worst-case Bob's beamforming matrix $\mathbf{w}_{\text{op}}$ and the antenna positions $\mathbf{t}_{\text{op}}$, under which the optimized secrecy rate, denoted by $R_{\text{op}}$, is then calculated. If $R_{\text{op}}$ is greater than $R_{\text{previous}}$, or if we have {a probability of $\rho$ that accepts the inferior solution.} then we update $\mathbf{w}_{\text{op}}$, $\mathbf{t}_{\text{op}}$ and $R_{\text{op}}$. Otherwise, we continue to maintain the current state and apply algorithms \ref{alg:PGA beamformer} and \ref{alg:PGA antenna position} to find better values. In addition, the probability of accepting the difference solution after each iteration also decreases, which can improve the effectiveness of the solution. The optimal beamforming matrix and MAs' position matrix are outputted when the maximum number of iterations is reached. {The above} method can reduce the possibility of falling into local optima during the process of optimization, then the optimal solution is found on a larger scale, and even the global optimal solution can be obtained. During the iterations, the influence of random variables can also lead to fluctuations of the secrecy rate. We can minimize this fluctuation by increasing the number of experiments. In particular, the simulations also demonstrate the effectiveness of our proposed method. Let \(N_a\) be the number of antennas being optimized. Thus, the complexity of algorithm \ref{alg:SA-PGA} scales with \(O(I_{\rm ter}^2 M_w + I_{\rm ter}^2 N_a M_t))\). }


\section{numerical results}\label{sec:evaluations}
{In this section, without the perfect CSI of the Eve, via MATLAB simulator, numerical results are provided to validate the effectiveness and correctness of our proposed algorithms for improving the secrecy performance of the MA-enabled secure communication system. In particular, on the one hand, we observe the impact of the number of moving MAs on the secrecy rate, under which we conclude that moving all MAs is not the best approach to maximize the secrecy rate. On the other hand, compared with two kinds of FPAs, namely UPA and ULA, the influence of the number of transmitting/receiving paths, path-loss exponent, noise power, and distance between the BS and the center of Eve's potential area. The BS transmits signals in 28GHz. In addition, the results are the average of 10000 ones.} 

\subsection{Secrecy Performance for Moving All and Parts of MAs}

\begin{figure}[htbp]
  \centering 
  \includegraphics[width=0.4\textwidth]{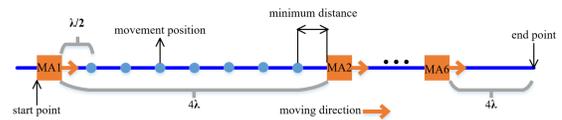}
  \caption{The structure of MA array based on ULA} 
  \label{fig:ULA array} 
\end{figure}

\begin{figure}[h]
  \centering 
  \includegraphics[width=0.35\textwidth]{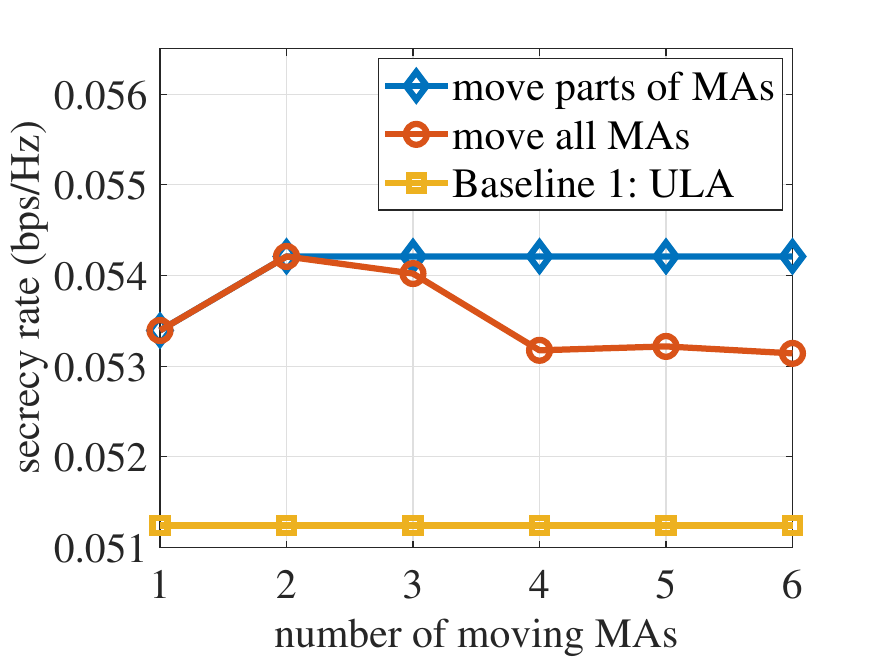}
  \caption{Secrecy performance for moving all and parts of MAs} 
  \label{fig:Verify hypotheses} 
\end{figure}

{Considering the structure of {the} MA array in Fig. \ref{fig:ULA array} consisting of 6 MAs, the minimum distance between any two MAs is set as $d_{\min}=\frac{\lambda}{2}$, and the moving range of each MA is $[0, 4\lambda]$ with the moving step size of $\frac{\lambda}{2}$, under which each MA can move 7 times. The transmit power of the BS is 10mW, the noise power is 0.5mW, and the length of potential area including an Eve is 2m. First, given a fixed beamforming matrix, we observe how the secrecy rate varies with different number of moving MAs. As shown in Fig. \ref{fig:Verify hypotheses}, via a one-dimensional search method, when moving all MAs, the secrecy rate shown by dashed red line first increases and then decreases. In fact, as described by dashed blue line, the secrecy rate can be maximized by only moving two MAs. For example, as shown in Table \ref{tab:Secrecy rate comparison of MA vs. FPA}, when moving the 3-th and 4-th MAs, it will cause a decrease in the secrecy rate, while moving the 5-th MA compensates for the decrement, but it is still worse than the optimal value. The reason behind the phenomenon is that the capability of only moving one antenna to increase the secrecy rate is limited. To sum up,  moving a part of MAs and fixing other ones not only simplifies the complexity of calculations but also improve the system efficiency. 
More important, a closer look suggests that the secrecy performance enhancement does not benefit unambiguously, which further emphasize the significance of joint design of BS's beamforming matrix and MAs's position. }

\begin{table}[htbp]
\centering
\caption{Secrecy rate comparison of MA vs. FPA}
\label{tab:Secrecy rate comparison of MA vs. FPA}
\setlength{\tabcolsep}{2.4mm}{
\begin{tabular}{*{6}{c}}
  \toprule
  \multirow{2}*{MA's number} & \multirow{2}*{ULA} & \multicolumn{2}{c}{move or not, vs. ULA} \\  
  \cmidrule(lr){3-4}
  & & moving parts of MAs & moving all MAs   \\
  \midrule
  1-th &0.0512& Yes, 4.195\% $\uparrow$& Yes, 4.195\% $\uparrow$\\
  2-th &0.0512& \textbf{\textcolor{red}{Yes, 5.787\%}} $\uparrow$& Yes, 5.787\% $\uparrow$\\
  3-th &0.0512& \textbf{No}, 5.787\% $\uparrow$& \textbf{Yes, 5.419\%} $\uparrow$  \\
  4-th &0.0512& \textbf{No}, 5.787\% $\uparrow$& \textbf{Yes, 3.768\%} $\uparrow$   \\
  5-th &0.0512& \textbf{No}, 5.787\% $\uparrow$& \textbf{Yes, 3.851\%} $\uparrow$ \\
  6-th &0.0512& \textbf{No}, 5.787\% $\uparrow$& \textbf{Yes, 3.702\%} $\uparrow$ \\
  \bottomrule
\end{tabular}}
\end{table}

\begin{figure}[htbp]
  \centering 
  \includegraphics[width=0.26\textwidth]{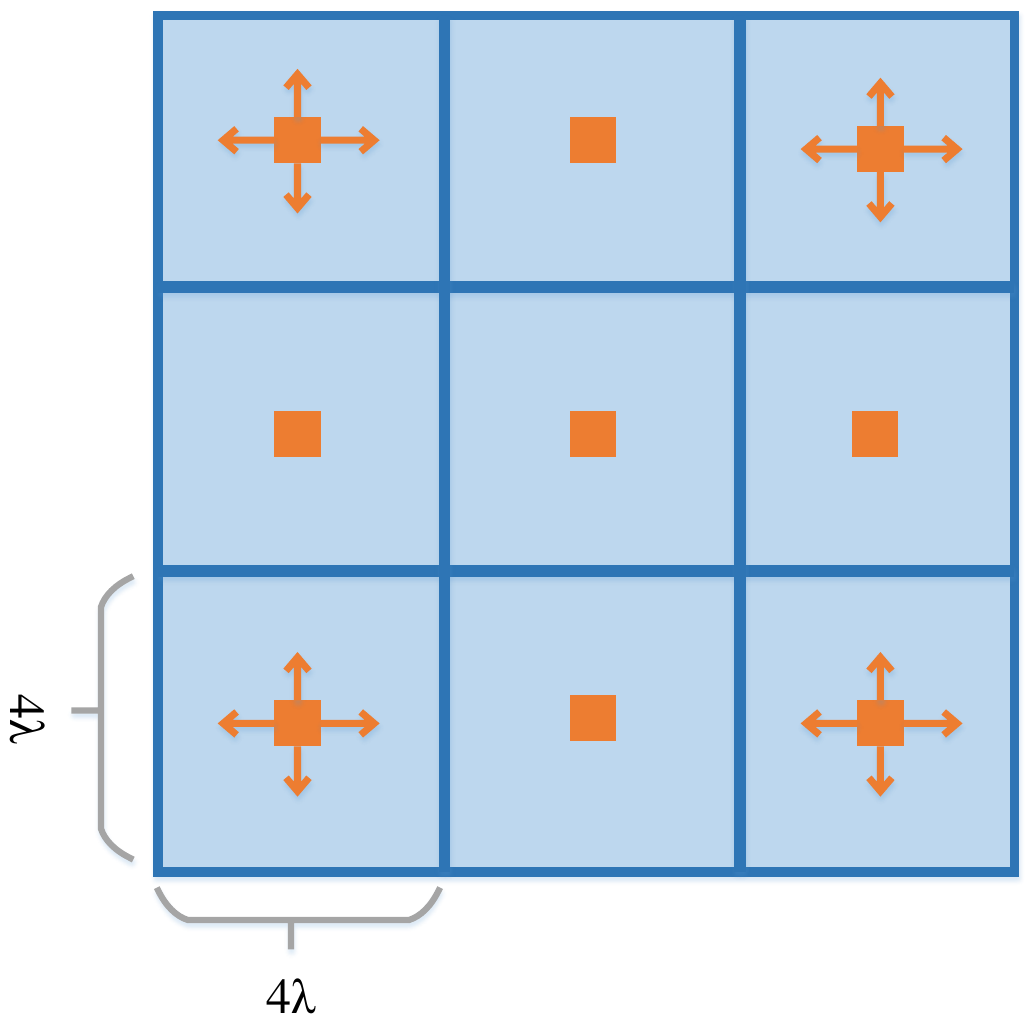}
  \caption{The structure of MA array based on UPA} 
  \label{fig:MA array} 
\end{figure}

{Next, based on the conclusion that not all MAs movements are beneficial for secrecy performance enhancement in Fig. \ref{fig:Verify hypotheses}, considering a structure of MA array in Fig. \ref{fig:MA array} consisting of uses a $3 \times 3$ UPA, and only MAs locating on the four corners move in the range of $[0, 4\lambda]$. Accordingly, the minimum distance between any two moving MAs is adjusted to $4\lambda$. In the following analyses, we validate the effectiveness of our proposed algorithms by moving 4MAs and fixing other 5 MAs.} 

\subsection{{Simulation Settings}}
{Considering 5 legitimate users, and the distance between them and the BS varies in the range of $[25, 35]$m, under which the positions of legitimate users are uniformly distributed. The length of square modeling the potential area of the Eve is 2m, and the distance between the BS and the center of potential area is 50m. In particular, regarding above the potential area of Eve as a UPA consisting of one virtual MA, and we model 3 potential positions of the Eve by virtual MA. The transmit power of the BS is 10mW, the noise power is 0.5mW, and other parameters are set according to Table \ref{table:parameter}.}

\begin{table}[h!]
\centering
\caption{Simulation Parameters}\label{table:parameter}
\begin{tabular}{lp{5cm}l} 
\toprule
Parameter & Description & Value \\
\midrule
\( \lambda \) & wavelength & 0.0107m \\
\( K \) & number of Bobs & 5 \\
\( M \) & number of virtual Eves & 3 \\
\( d_t \) & distance between the Bobs and the BS & [25,35]m \\
\( d \) & distance between the center of Eve's potential area and the BS & 50m \\
\( r \) & length of Eve's potential area & 2m \\
\( P_t \) & transmit power & 10mW \\
\( P_{\text{noise}} \) & noise power & 0.5mW \\
\( L \) & number of paths & 3 \\
\( g_0 \) & average channel gain at reference of 1m & 30dB \\
\( \alpha \) & path-loss exponent & 2 \\
\( N \) & number of MAs & 9 \\
\( A \) & movement range of each MA & \( 4\lambda \) \\
\( T \) & Initial Temperature of Simulated Annealing & 1 \\
\( \beta \) & cooling coefficient of simulated annealing & 0.9 \\
\( \delta_w \) & step size when optimizing $\textbf{W}$ & \( 0.01 \) \\
\( N_{\text{sample}} \) & number of sampled generalization & 10 \\
\( \delta_T \) & step size when optimizing $\textbf{T}$ & \( 0.001 \) \\
\( \tau_w \) & optimization threshold comparison for \( \mathbf{w} \) & \( 0.005\) \\
\( \tau_T \) & optimization threshold comparison for \( \mathbf{T} \) & \( 0.0001 \) \\
\( I_{\rm ter} \) & maximum number of iterations & 1000 \\
\bottomrule
\end{tabular}
\label{tab:simulation_parameters}
\end{table}

{For the channel between the BS and the Bobs/Eve, 
geometric channel models are used since they accurately represent the one-to-one correspondence between the transmission and reception paths \cite{ref19}, under which the path response matrix is a diagonal matrix, where diagonal elements $\sigma$ follow a complex normal distribution with the mean of zero and variance of $\frac{g_0 }{L}d^{-\alpha}$, where $g_0$ is the average channel power gain at a reference distance of 1m, $L$ is the number of transmitting/receiving paths, $d$ is the distance between the BS and the Bobs/Eve, and $\alpha$ is the path-loss exponent. 
In detail, both the number of transmitting and receiving paths are set as 3, $g_0=30\text{dB}$, and $\alpha=2$, respectively.}

\subsection{Convergence Evaluation of Proposed SA-PGA Method}
\begin{figure}[htbp]
  \centering 
  \includegraphics[width=0.4\textwidth]{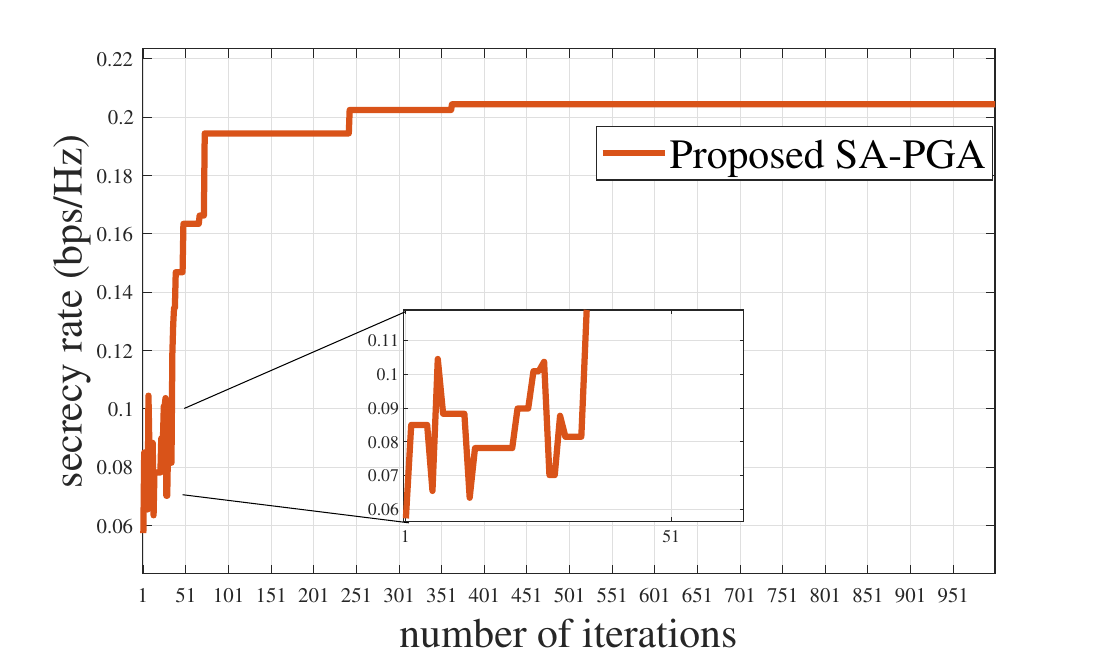}
  \caption{Convergence of secrecy rate obtained by SA-PGA} 
  \label{fig:Iterative curve} 
\end{figure}

{Monte Carlo simulations are employed to approximate the gradient, because of the random variables in gradient computation. 
Furthermore, due to the non-convex nature of the objective function with respect to two optimization variables, the AdaGrad algorithm with an initialize adaptive step size of 0.01 and 0.001 for $\mathbf{w}_{k^*}$ and $\mathbf{t}_n$ is used to enhance the convergence of the algorithm during the optimization of beamforming and MAs' positions. 
In addition, for similar reasons, the SA-PGA method may still fall into locally optimal solution or exhibit fluctuations in the objective function values after optimization. Therefore, the simulated annealing algorithm is integrated to search for the global optimum. The main idea is to compute the secrecy rate after optimizing the beamforming and MAs' positions. If the calculated secrecy rate is lower than the value obtained in the previous iteration, the algorithm reverts to the settings of beamforming and MAs' positions from the previous optimization and conducts another search. Alternatively, a worse solution may be accepted with a small probability, which decreases as iterations progress, allowing for exploration of a broader space in search of higher-quality solutions. Fig. \ref{fig:Iterative curve} shows the convergence curve of the secrecy rate over the number of iterations. }

{From Fig. \ref{fig:Iterative curve}, it can be seen that before the initial stage when the number of iterations is less than or equals to 35 times, a significant fluctuation in the secrecy rate exists, and low-quality solutions are accepted to obtain a more extensive search in the solution space. When the number of iterations varies from 35 to 364 times, the proposed SA-PGA method quickly {reaches} a relatively stable stage, which may indicate finding a local or global optimal solution and decreasing the probability of accepting inferior solutions, with almost no acceptance of inferior solutions. After 364 iterations, the secrecy rate remains unchanged, which means that the algorithm  has converged to an optimal solution. To sum up, It can be concluded that the proposed SA-PGA method can effectively solving optimization problems. 
}
\begin{figure}[htbp]
\centering  
\subfigure[Original antenna position]{
\includegraphics[width=0.23\textwidth]{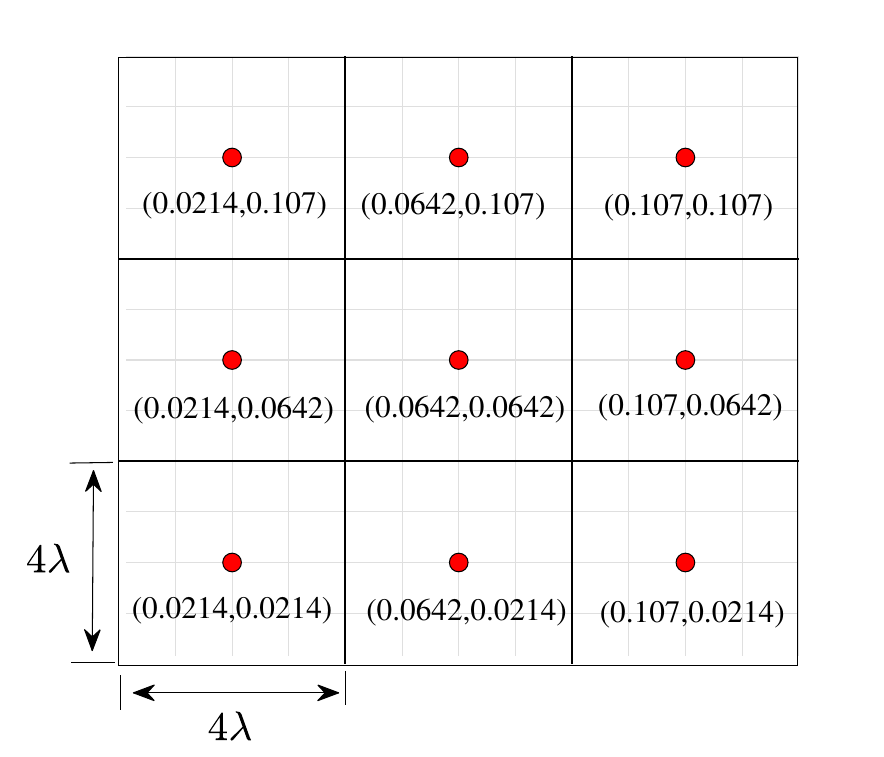}\label{sub:antenna 1}}
\subfigure[Optimal antenna position]{
\includegraphics[width=0.227\textwidth]{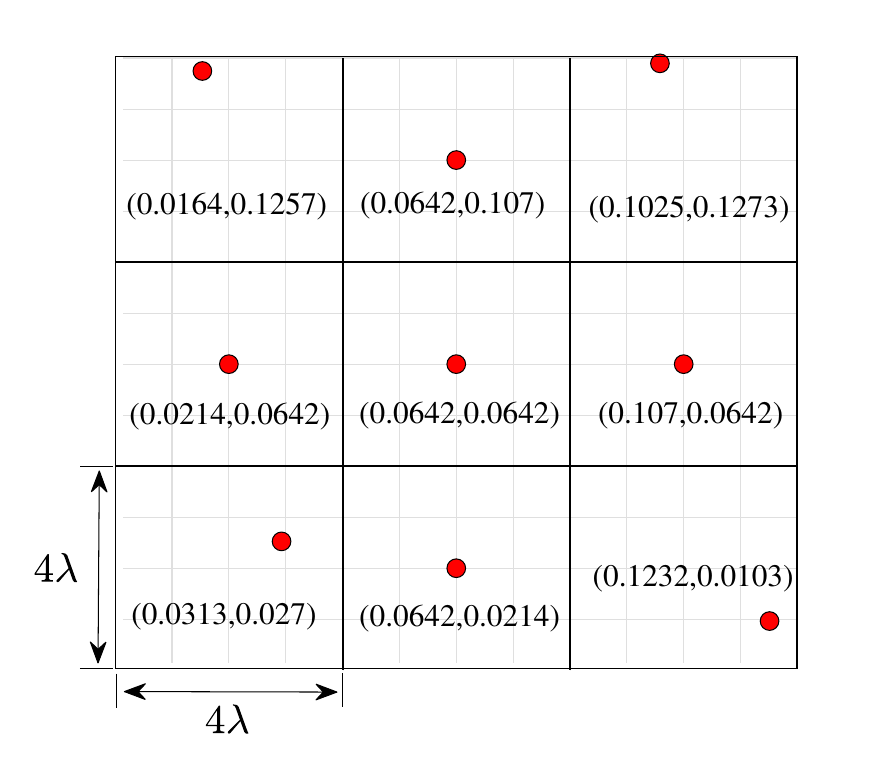}\label{sub:antenna 2}}
\caption{Position comparison of Antenna array}
\label{fig:Position comparison of Antenna array} 
\end{figure}

{When the secrecy rate achieved by joint design of beamforming and MA's positions reaches the convergence, the corresponding positions of all MAs is shown in Fig. \ref{sub:antenna 2}. For comparison, the initial positions of all MAs is shown in Fig. \ref{sub:antenna 1}. Then, given the optimal beamforming and MAs' positions, we validate the impact of transmitting/receiving paths, path-loss exponent, noise power, and distance between the BS and the center of Eve's potential area on secrecy rate.}

\subsection{Secrecy Performance Comparison with FPA }
{In this subsection, we compare the secrecy rate obtained by joint design of beamforming matrix and MA's positions with these of ULA and UPA. In particular, both ULA and UPA adopt a field response channel model.    }

\begin{figure}[htbp]
  \centering 
  \includegraphics[width=0.35\textwidth]{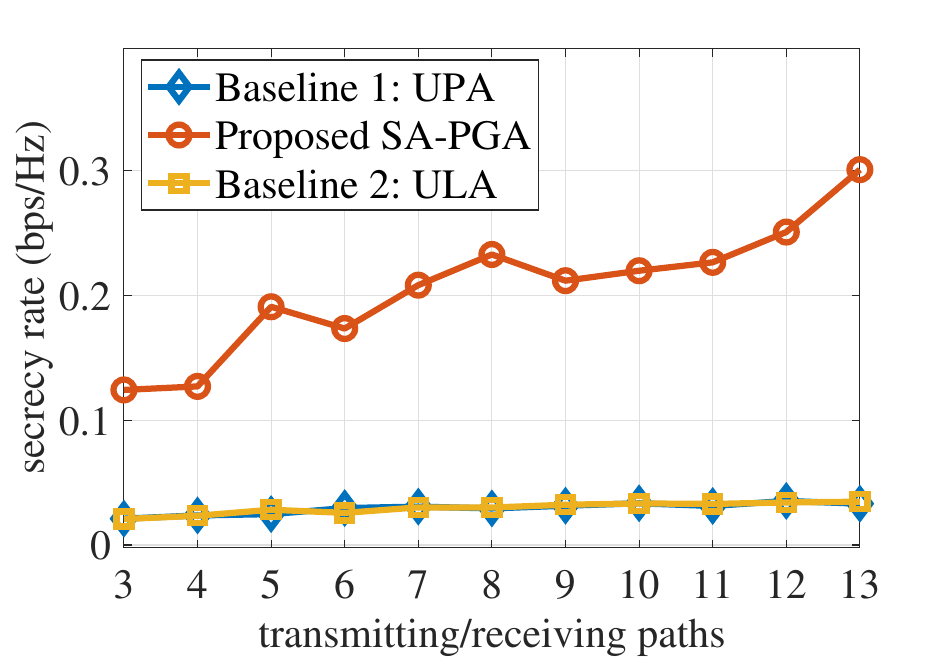}
  \caption{Secrecy rate vs. number of paths} 
  \label{fig:sec_path} 
\end{figure}

{Using parameter settings in Table \ref{tab:simulation_parameters}, we observe the influence of the number of transmitting/receiving paths on the secrecy rate, as shown in Fig. \ref{fig:sec_path}. It can be seen that the secrecy rate obtained by three kinds of antennas performs a upward trend. In detail, the secrecy rate achieved by ULA and UPA increases slowly with the number of paths, and the performance difference between the two kinds of antennas is not significant. One possible reason is that increasing the number of paths does not effectively enhances legitimate channel capacity or suppresses eavesdropping channel capacity due to the introduced randomness. The secrecy rate of proposed SA-PGA increases significantly, and gives the best secrecy performance for all the number of paths. The reason behind this phenomenon is that the number of paths can directly affect the dimension of the path response matrix, and a larger number of paths will increase the randomness of the channel. Therefore, we can conclude that the increased randomness of the path is helpful for the secrecy performance enhancement.} 

\begin{figure}[htbp]
\centering  
\subfigure[secrecy rate ]{
\includegraphics[width=0.33\textwidth]{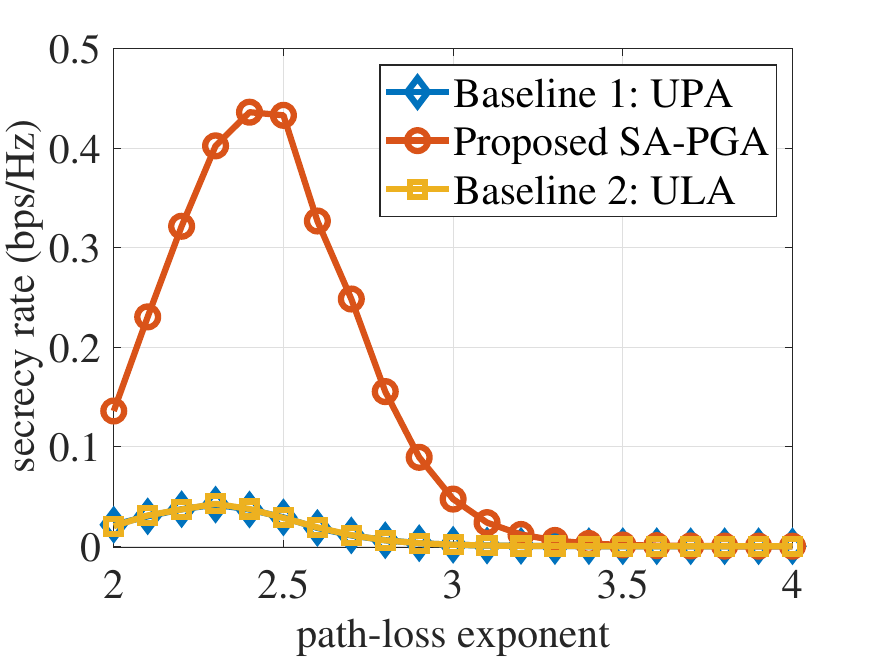}\label{sub:path-loss 1}}
\subfigure[channel capacity of Bob]{
\includegraphics[width=0.23\textwidth]{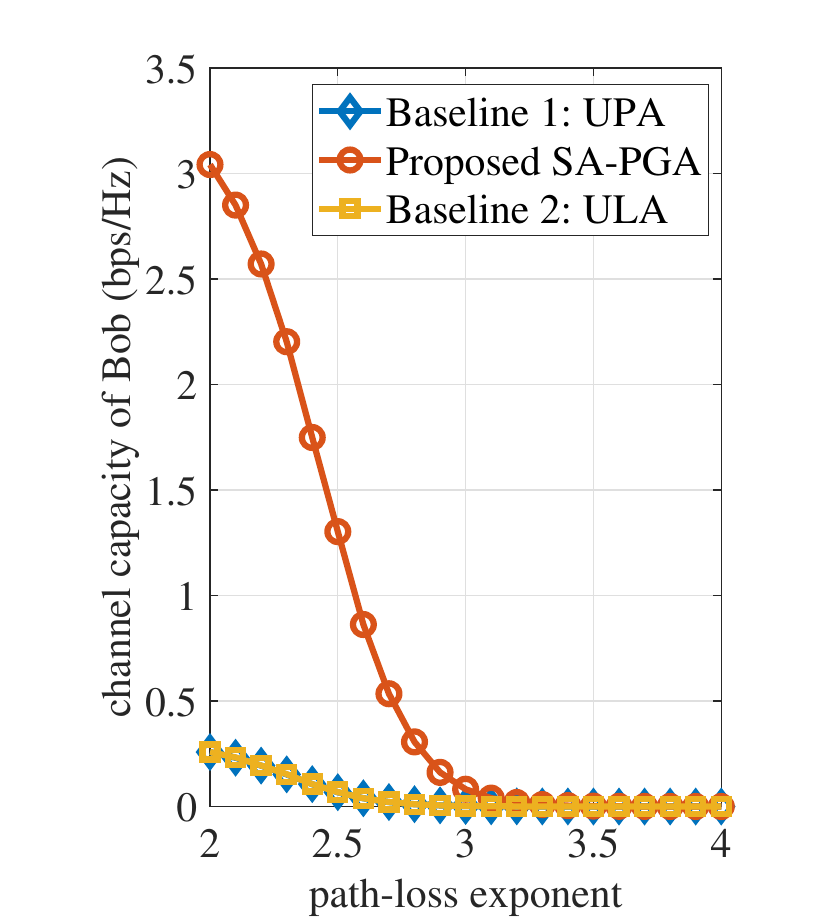}\label{sub:path-loss 2}}
\subfigure[channel capacity of Eve]{
\includegraphics[width=0.23\textwidth]{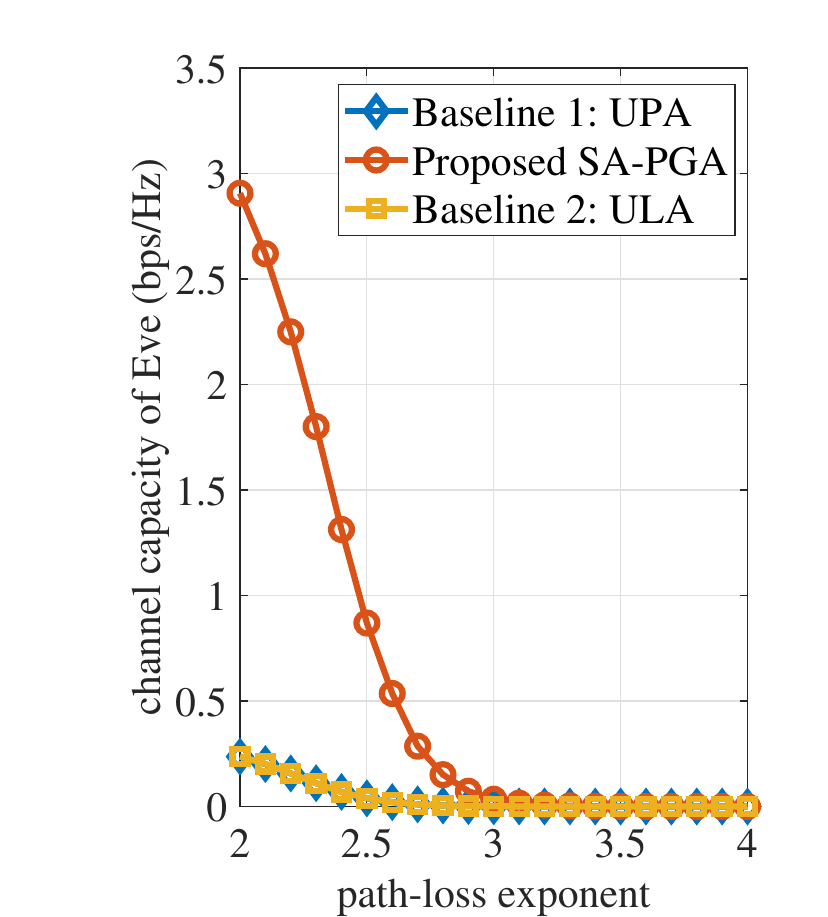}\label{sub:path-loss 3}}
\caption{Channel capacity and secrecy rate for different settings of path-loss exponent}
\label{fig:sec_a} 
\end{figure}

{Next, we consider the impact of the path-loss exponent on the secrecy rate, other parameters are set based on Table \ref{tab:simulation_parameters}, as shown in Fig. \ref{fig:sec_a}.
In particular, it can be seen that there is a non-linear relationship between the secrecy rate and path-loss exponent among the three methods. The secrecy rate shows a rapid upward trend when $\alpha$ varies from 2 to 2.4, which is followed by a rapid decrease in the secrecy rate after $\alpha=2.4$, and reaches a stable convergence when $\alpha$ is greater than 3.3. Similarly, ULA and UPA show an upward trend when $\alpha$ varies from 2 to 2.3, which is followed by a decrease and a stable convergence after $\alpha=2.9$. 
The reason for this phenomenon is that by carefully comparing the channel capacities of the Bob and Eve, as shown in Fig. \ref{sub:path-loss 2} and Fig. \ref{sub:path-loss 3} respectively, it can be observed that the channel capacity of the eavesdropping channel is more sensitive to changes when $\alpha$ is less than about 2.4. Accordingly, the channel capacity of the eavesdropping channel decreases faster than that of Bobs, resulting in an increase in the secrecy rate. Afterwards, the channel capacity of Bobs decreases faster than that of Eve, resulting in a decrease in secrecy rate. The corresponding reason that the secrecy rates obtained by UPA and ULA first increases and then decreases are similar to that of the proposed method. In addition, the secrecy rate obtained by the SA-PGA method shows the best secrecy performance among three kinds of antennas, and reflects a good environmental adaptability for all different settings of the path-loss exponent.}

\begin{figure}[htbp]
\centering  
\subfigure[secrecy rate ]{
\includegraphics[width=0.35\textwidth]{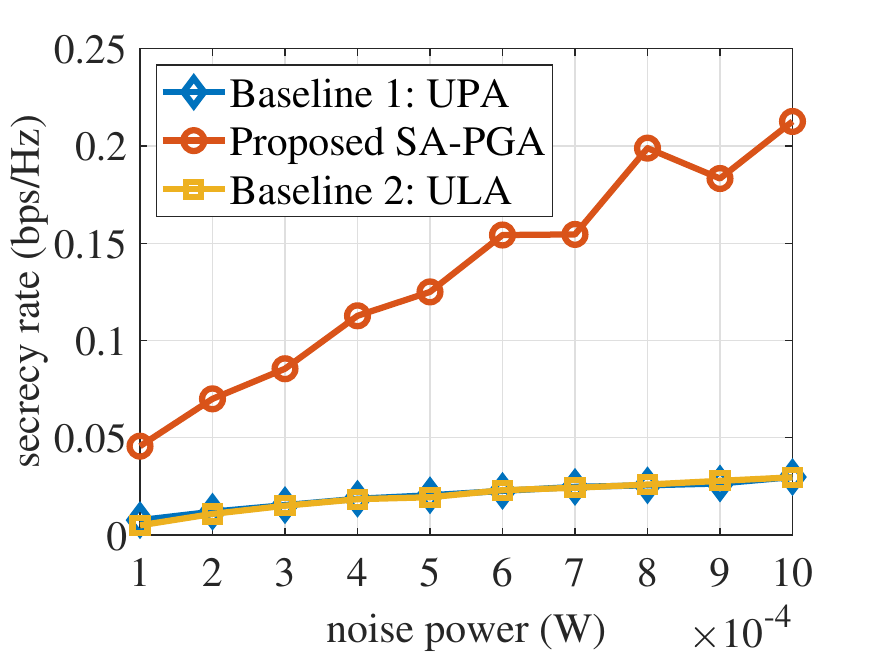}\label{sub:noise 1}}
\subfigure[channel capacity of Bob]{
\includegraphics[width=0.23\textwidth]{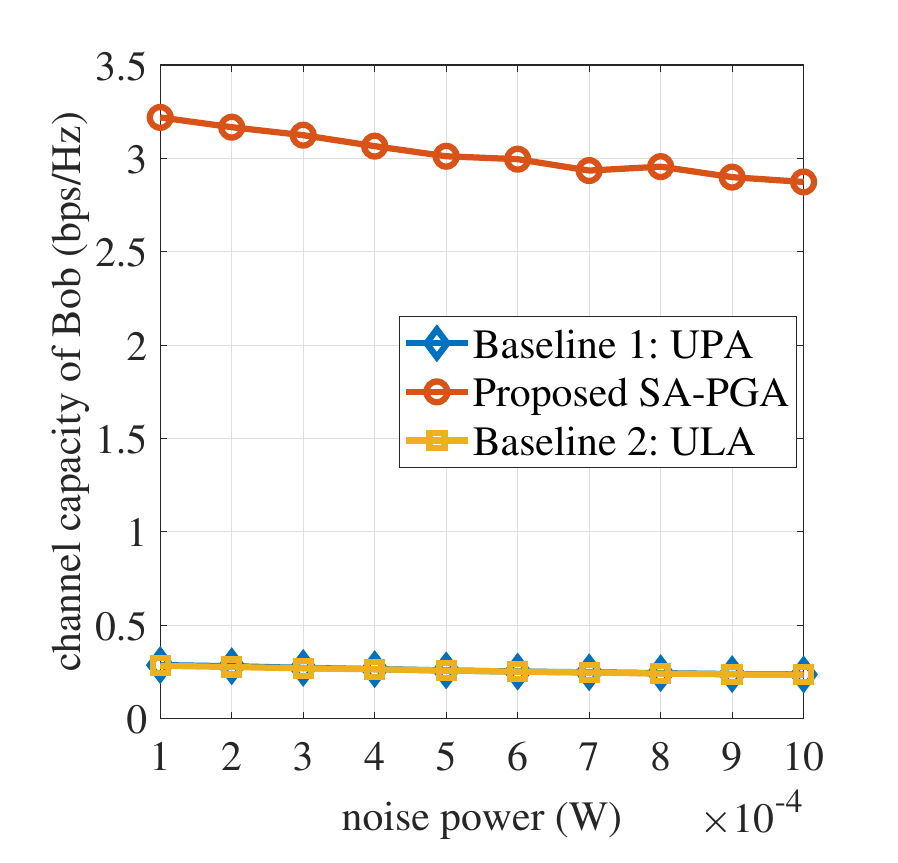}\label{sub:noise 2}}
\subfigure[channel capacity of Eve]{
\includegraphics[width=0.23\textwidth]{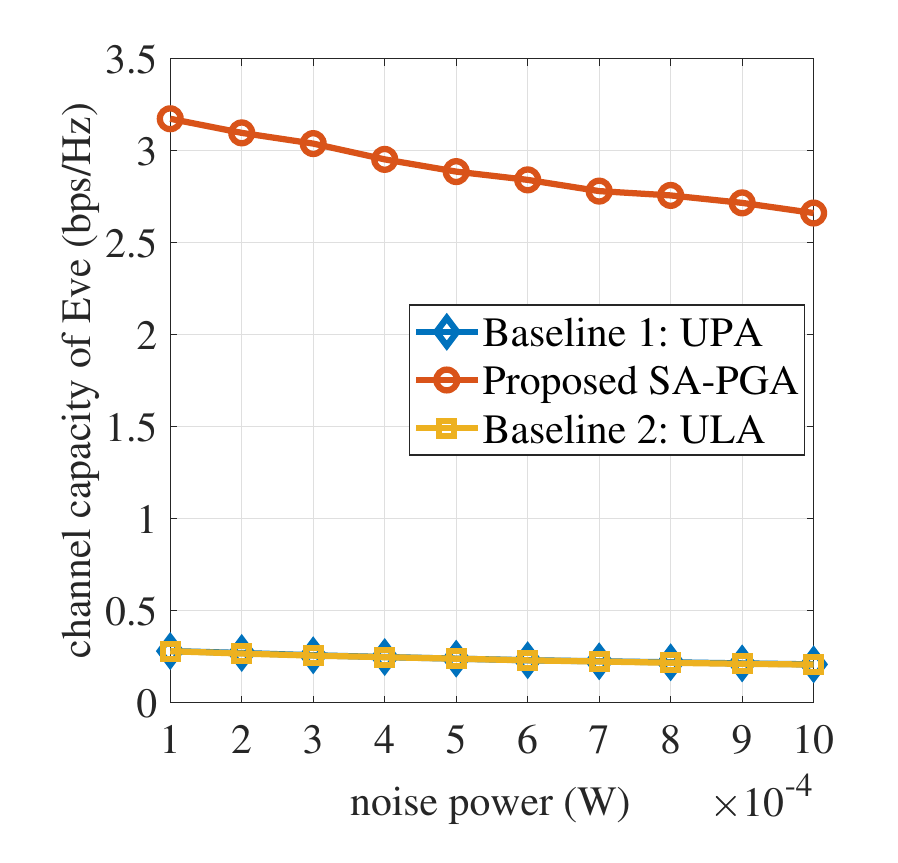}\label{sub:noise 3}}
\caption{Channel capacity and secrecy rate for different settings of noise power}
\label{fig:sec_noise} 
\end{figure}

{Finally, the influence of the path-loss exponent on the secrecy rate is given in Fig. \ref{fig:sec_noise}. It can be seen that the
secrecy rate obtained by three kinds of antennas performs a upward trend. In detail, the secrecy rate achieved by ULA and UPA increases slowly with the noise power, and the performance difference between the two is not significant. One possible reason is that in the presence of a response channel model, there is no essential difference between ULA and UPA. The secrecy rate of the proposed SA-PGA method increases significantly, and gives the best secrecy  performance for all settings of the noise power. The reason is that, 
from Fig. \ref{sub:noise 2} and Fig. \ref{sub:noise 3}, it can be seen that the capacity of eavesdropping channel decreases faster, which means that it is more sensitive to the noise power, and thus the obtained secrecy rate gradually increases.  In this way, adding some noise power is an available method to effectively suppress the eavesdropping channel, thereby increasing the system security.}

\section{Conclusions}\label{sec:conclusion}
In this paper, we leveraged MAs technology to enable PLS {without} Eve's CSI. By jointly optimizing the beamforming vector and antenna positions of all MAs at Alice, we constructed an optimization problem to maximize the secrecy rate of the worst-case Bob. The non-convex optimization problem was effectively addressed by using the SA-PGA method. A performance comparison reveals that MA-based PLS significantly enhanced the security performance compared to the conventional FPA ones.

\end{document}